\documentclass[a4paper,fleqn]{cas-dc}

\usepackage{cmap} 
\usepackage[T1]{fontenc}
\usepackage{lmodern}
\usepackage[numbers]{natbib}
\usepackage{graphicx}
\usepackage{amsmath}
\usepackage{amssymb}
\AtBeginDocument{\let\sim\thicksim}
\usepackage{booktabs}
\usepackage{url}
\usepackage{hyperref}
\usepackage{float}
\usepackage{caption}
\usepackage{subcaption}
\usepackage{stfloats}
\usepackage{cuted}
\usepackage{enumitem}
\usepackage{multirow}
\usepackage{xcolor}
\usepackage{colortbl}
\usepackage{tikz}
\usetikzlibrary{shapes.geometric, arrows.meta, positioning, calc, fit}
\usepackage{caption} 
\usepackage[left]{lineno}

\def\tsc#1{\csdef{#1}{\textsc{\lowercase{#1}}\xspace}}
\tsc{WGM}
\tsc{QE}

\begin{document}
\let\WriteBookmarks\relax
\def\floatpagepagefraction{1}
\def\textpagefraction{.001}

\shorttitle{Visual-First Multimodal RAG for Civil Standard Plans}

\shortauthors{N. Subedi et al.}

\title [mode = title]{PlanSightRAG: A Visual-First Multimodal RAG for  Automating Question Answering and Compliance Checking for Civil Standard Plans}

\author[1]{Nabaraj Subedi}[orcid=0009-0005-1596-8641]
\ead{nsubedi1@uwyo.edu}
\credit{Conceptualization, Methodology, Software, Writing -- Original Draft}

\author[2]{Shuvo Dip Datta}[orcid=0000-0001-8604-3148]
\ead{sdatta1@uwyo.edu}
\credit{Investigation (literature review), Writing -- Original Draft}

\author[2]{Ahmed Abdelaty}[orcid=0000-0001-7329-452X]
\cormark[1]
\ead{aahmed3@uwyo.edu}
\credit{Conceptualization, Supervision, Review \& Editing}

\author[1]{Shivanand Venkanna Sheshappanavar}[orcid=0000-0003-4039-2910]
\ead{ssheshap@uwyo.edu}
\credit{Supervision, Review \& Editing}

\affiliation[1]{organization={Department of Electrical Engineering \& Computer Science, University of Wyoming}}

\affiliation[2]{organization={Department of Civil \& Architectural Engineering \& Construction Management, University of Wyoming},
            addressline={1000 E. University Ave.},
            city={Laramie},
            postcode={82071},
            state={WY},
            country={USA}}

\cortext[1]{Corresponding author}

\begin{abstract}
Civil infrastructure compliance checking has long relied on engineers manually reading legacy 2D plans; however, OCR-based automation strips away the geometry and layout essential for interpreting these plans. We present a Visual-First Multimodal Retrieval-Augmented Generation (RAG) framework called \textbf{PlanSightRAG}. It indexes and reasons directly over plan imagery, integrates a ColNomic-3B multi-vector retrieval, an agentic Planner–Retriever–Auditor–Synthesizer, and MaxSim heatmaps as an evidence trail. We introduce a 4,056-pair benchmark from five state Departments of Transportation (DOT) standard plans (1,898-pages). PlanSightRAG achieves \textbf{91.47\%} Recall@5 on zero-shot retrieval, while on a held-out Michigan DOT corpus, it achieves \textbf{91.40\%}. On synthetic, parametrically-generated compliance drawings, our Qwen2.5-VL-72B pipeline reaches \textbf{100\%} verdict accuracy \emph{only when supplied a pre-resolved rule threshold}, a controlled ceiling that a non-VLM OCR baseline already reaches at 76.4\%. Finally, we demonstrate autonomous visual rule-grounding by extracting numeric limits directly from a specification corpus without any human-supplied rules.
\end{abstract}


\begin{highlights}
\item Visual-first multimodal RAG audits civil standard plans without OCR.
\item ColNomic-3B reaches 91.47\% Recall@5 on a new 4,056-pair five-DOT benchmark.
\item Zero-shot retrieval transfers to an unseen DOT at 91.40\%; LoRA tuning adds nothing.
\item Agentic pipeline: 100\% verdicts with resolved thresholds; transfers to real plans.
\item First visual rule-grounding: numeric limits extracted with no human-supplied rule.
\end{highlights}


\maketitle

\section{Introduction}\label{sec:introduction}

Compliance checking underpins the safety and long-term reliability of civil infrastructure, yet in most state transportation agencies, it is still carried out by engineers reading through dense 2D standard plans. The plans are not obsolete—DOTs continue to maintain thousands of legacy sheets that remain the authoritative reference for design review—and the workflow is expensive, slow, and error-prone on multi-sheet sets with complex spatial conventions \citep{picard2025concept}. A way to query and audit these archives at scale while keeping humans in the loop is missing.

RAG and Document Question Answering frameworks have been natural candidates \citep{lewis2020retrieval,borgeaud2022improving,izacard2023atlas,ram2023context}, in which documents are linearized via OCR into tokens that an embedding model scores. On engineering plans, these embeddings lack geometric properties, positional layout, and cross-view symbols; hence, they lose the primary semantics of a standard plan. Once the standard plans are reduced to a token bag, the retriever can no longer locate the relevant plan reliably \citep{xu2020layoutlm}. We therefore treat this as the central gap and propose a multi-vector retrieval framework in which the retrieval layer itself sees the plan. Figure~\ref{fig:intro_comparison} makes this gap concrete on a realistic WYDOT query. 

\bigskip
\noindent
\begin{minipage}{\linewidth}
\centering
\resizebox{\linewidth}{!}{\begin{tikzpicture}[
  >=Latex, font=\sffamily,
  querybox/.style={draw=purple!60, fill=purple!5, rounded corners, align=center,
                   inner sep=4pt, text width=8.2cm, font=\sffamily\footnotesize}, 
  rowbox/.style={draw=#1!60!black, fill=#1!5, rounded corners,
                 minimum width=8.6cm, minimum height=1.9cm, inner sep=0pt,
                 anchor=north},
  rowlabel/.style={font=\sffamily\bfseries\footnotesize, text=#1!50!black,
                   anchor=north west},
  rowdesc/.style={font=\sffamily\scriptsize, anchor=north east, align=right,
                  text width=4.8cm},
  ocrtoken/.style={font=\sffamily\tiny\ttfamily, draw=red!40!black,
                   fill=red!10, rounded corners=0.15ex, inner sep=1pt},
]

\node[querybox, anchor=north] (q) at (0, 0)
  {\textbf{Query:} \textit{stake spacing \& wire-tie config for wire-enclosed riprap (WYDOT 511-1A)?}};

\node[rowbox=red] (r1) at (0, -1.2) {};

\node[rowlabel=red] at ([xshift=5pt,yshift=-4pt]r1.north west)
  {OCR $+$ BM25 Text RAG \textcolor{red!60!black}{\large$\times$} FAILS};

\begin{scope}[shift={([xshift=5pt,yshift=8pt]r1.south west)}]
  \node[ocrtoken] at (0.35, 0.55) {WIRE};
  \node[ocrtoken] at (1.15, 0.70) {TYP};
  \node[ocrtoken] at (2.05, 0.55) {SPACING};
  \node[ocrtoken] at (0.50, 0.15) {FT};
  \node[ocrtoken] at (1.30, 0.25) {SEE};
  \node[ocrtoken] at (2.30, 0.15) {RIPRAP};
\end{scope}

\node[rowdesc] at ([xshift=-6pt,yshift=-22pt]r1.north east)
  {Token bag $\Rightarrow$ geometry lost\\[2pt]
   \textbf{\textcolor{red!60!black}{Recall@5: 58.02\%}}};

\node[rowbox=orange] (r2) at (0, -3.25) {};

\node[rowlabel=orange] at ([xshift=5pt,yshift=-4pt]r2.north west)
  {VisionRAG (Pyramid) \textcolor{orange!60!black}{\large$\times$} FAILS};

\begin{scope}[shift={([xshift=5pt,yshift=8pt]r2.south west)}]
  \draw[fill=orange!15, draw=orange!60!black, rounded corners=0.2ex]
        (0,0) rectangle (2.6,0.9);
  \foreach \y in {0.15,0.35,0.55,0.75}{
    \draw[orange!40!black, line width=0.9pt, opacity=0.55] (0.15,\y)--(2.45,\y);
  }
  \node[font=\tiny\sffamily\itshape, text=orange!70!black] at (1.3,0.45) {blurred};
\end{scope}

\node[rowdesc] at ([xshift=-6pt,yshift=-22pt]r2.north east)
  {Low-res $\Rightarrow$ fine details lost\\[2pt]
   \textbf{\textcolor{orange!70!black}{Recall@5: 45.99\%}}};

\node[rowbox=green!50!black] (r3) at (0, -5.3) {};

\node[rowlabel=green!45!black] at ([xshift=5pt,yshift=-4pt]r3.north west)
  {ColNomic (Ours) \textcolor{green!45!black}{\large$\checkmark$} SUCCEEDS};

\begin{scope}[shift={([xshift=5pt,yshift=8pt]r3.south west)}]
  \draw[fill=green!8, draw=green!55!black, rounded corners=0.15ex]
        (0,0) rectangle (2.6,0.9);
  \foreach \x in {0.26,0.52,...,2.34}{
    \draw[green!50!black, line width=0.25pt] (\x,0)--(\x,0.9);
  }
  \foreach \y in {0.15,0.3,0.45,0.6,0.75}{
    \draw[green!50!black, line width=0.25pt] (0,\y)--(2.6,\y);
  }
  \node[font=\tiny\sffamily\bfseries, text=green!50!black]
       at (1.3,0.45) {$32{\times}32$};
\end{scope}

\node[rowdesc] at ([xshift=-6pt,yshift=-22pt]r3.north east)
  {Multi-vector MaxSim $\Rightarrow$ grounded\\[2pt]
   \textbf{\textcolor{green!45!black}{Recall@5: 92.69\%}}};

\end{tikzpicture}}

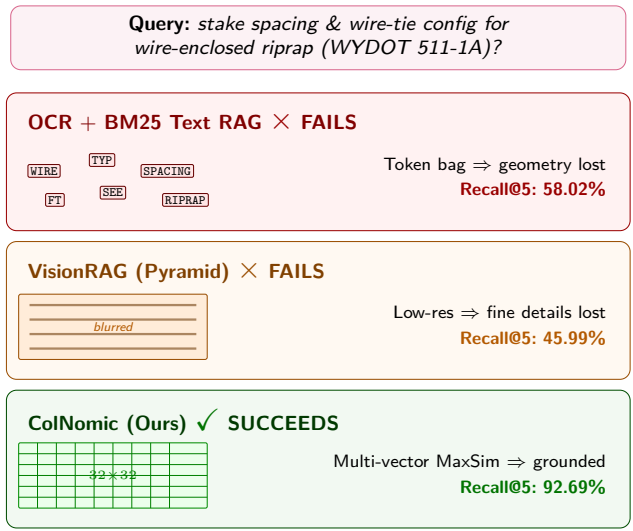
\captionof{figure}{Existing retrieval failure cases and patch-level visual retrieval success on a realistic WYDOT query.}\label{fig:intro_comparison}
\end{minipage}
\newline

Two recent lines of work make a visual-first approach plausible. The Vision-centric retrieval model ColPali~\citep{faysse2025colpali} indexes a page as a grid of patch embeddings and scores queries via late-interaction MaxSim, preserving layout without text conversion. In parallel, vision-language models (VLMs), including Qwen-VL~\citep{bai2023qwenvl} and Gemini~\citep{team2025gemini}, have demonstrated fine-grained visual grounding on diagrams. These systems are usually evaluated in isolation; however, neither the retrieval nor the reasoning layer has been integrated into a pipeline capable of auditing multi-sheet engineering archives end-to-end. As Figure~\ref{fig:intro_comparison} shows, a text-based RAG pipeline and the hybrid VisionRAG baseline both miss the dimension chain and wire-tie detail. On the contrary, the patch-level visual indexing retrieves the correct standard plan at Top-1 and grounds its answer on the relevant region (Table~\ref{tbl:comparison} reports the full quantitative comparison). In this paper, we close the gap with a Visual-First Multimodal RAG framework called \textbf{PlanSightRAG} that indexes plans directly as images, reasons over the retrieved plans with a VLM, and decomposes multi-plan compliance queries through a dedicated agentic pipeline. The central contribution is a single end-to-end, OCR-free system for \emph{automated compliance checking} of civil standard plans; visual retrieval, MaxSim grounding, agentic auditing, and autonomous rule-grounding are the system components, and the five-DOT benchmark is the substrate on which we validate it. Our contributions are threefold:

\begin{enumerate}
     \item \textbf{Visual-First Retrieval Framework and Five-DOT Benchmark:} A multi-vector late-interaction (ColNomic-3B) visual indexing pipeline with high-resolution tiling to retrieve engineering plans without OCR.

     \item \textbf{Agentic Grounded Compliance Pipeline:} A Planner–Retriever–Auditor–Synthesizer pipeline with sharpened MaxSim heatmaps that audit designs against retrieved standards via a transparent evidence trail.

    \item \textbf{Autonomous Visual Rule-Grounding:} Instead of hand-injecting the rule, the agent retrieves the governing requirement from a specification corpus (Recall@5 $=100\%$ among 1,913 candidates), extracts its numeric limit (resolving symbolic $d/2$), and audits against the \emph{self-grounded} threshold, matching the 100\% ceiling with no human-supplied rule.
\end{enumerate}


We ground the evaluation in four pre-declared, falsifiable hypotheses (\textbf{H1}--\textbf{H4}) rather than qualitative claims, so that each finding can be read as supported, partially supported, or rejected against an explicit numeric threshold declared with each test. \textbf{H1} asks whether patch-level visual retrieval surpasses OCR-based and hybrid alternatives. \textbf{H2} asks whether the agentic Planner--Retriever--Auditor--Synthesizer pipeline, armed with per-drawing rule thresholds, audits dense multi-view sheets and multi-plan designs. \textbf{H3} concerns domain adaptation: whether LoRA fine-tuning of the adopted ColNomic-3B retriever on the five-DOT training split improves held-out Recall@5 without degrading zero-shot transfer to the unseen Michigan DOT corpus (Section~\ref{sec:res-retrieval-ablation}). Finally, \textbf{H4} tests whether high-resolution tiling improves judge accuracy.




\section{Literature Review}\label{sec:litreview}

\subsection{Automated Compliance Checking in Civil Infrastructure}\label{sec:lit-compliance}

Automated compliance checking has traditionally relied on rule-based and text-driven approaches that encode regulatory requirements as machine-executable logic \citep{eastman2009automatic,solihin2015classification}. These methods fundamentally assume that compliance information can be represented as structured data or linear text. However, for legacy 2D standard plans, requirements are encoded implicitly through geometry, layout, and cross-view symbols, encouraging an error-prone manual visual inspection as the dominant current practice \citep{tang2010automatic}.

\subsection{OCR-Based Document QA and Text-Centric RAG Systems}\label{sec:lit-ocr}

RAG \citep{lewis2020retrieval} grounds language models in external knowledge and has been deployed successfully for regulatory clause retrieval and general-purpose Document Question Answering (QA) \citep{xu2020layoutlm,xu2021layoutlmv2,appalaraju2021docformer,wu2022natural,mathew2021docvqa}, where rules and answers are explicitly stated in the text. These successes are the reference point against which visual-first retrieval must be compared; the failure modes that motivate this paper have already been discussed in Section~\ref{sec:introduction}. What is most relevant for positioning is that layout-aware text encoders \citep{xu2021layoutlmv2,appalaraju2021docformer} and image-to-structure approaches \citep{roberts2024image2struct,liu2023matcha} recover part of the spatial signal lost during OCR. However, they do not fully restore this information. Recent diagnostic evaluations further indicate that the remaining spatial loss is sufficient to degrade engineering-compliance tasks \citep{picard2025concept}.

\subsection{VLMs for Drawing Understanding}\label{sec:lit-vlm}

While early VLMs primarily targeted natural images and explicitly structured documents \citep{antol2015vqa,radford2021learning}, recent advances have yielded state-of-the-art (SOTA) transformer-based models capable of processing high-resolution visual contexts. Some of the VLMs include robust open-source foundations such as Qwen-VL \citep{bai2023qwenvl}, InternVL \citep{chen2024internvl,wang2025internvl35}, BLIP-2 \citep{li2023blip2}, Flamingo \citep{alayrac2022flamingo}, and LLaVA \citep{liu2024visual}, as well as closed-source alternatives like Gemini \citep{team2025gemini} and GPT-4V \citep{openai2023gpt4,achiam2024gpt4v}. These models achieve strong performance on general multimodal benchmarks; however, their application directly to dense, domain-specific technical documents remains underexplored.

In the civil engineering and construction domain, several studies have explored VLM-based approaches for technical document understanding \citep{picard2025concept,shteriyanov2025blueprintsymvl}. However, multiple studies emphasize that generic VLMs face significant limitations when applied to engineering drawings \citep{picard2025concept,techmb2025}. OCR-centric or region-based pipelines often fail to capture implicit relationships encoded through layout, geometry, symbols, and cross-view references, which are central to technical drawings and standard plans \citep{shteriyanov2025blueprintsymvl}. Chart and diagram understanding benchmarks such as PlotQA \citep{methani2020plotqa}, FigureQA \citep{kahou2018figureqa}, ChartQA \citep{masry2022chartqa}, and DVQA \citep{kafle2018dvqa} further show the limits of current VLMs. Their performance drops on tasks that require precise spatial grounding, symbolic interpretation, or multi-step visual reasoning, even when textual annotations are available. Recent work on understanding engineering drawings confirms that general-purpose VLMs struggle with the visual conventions unique to technical drafting. The HallusionBench diagnostic suite \citep{guan2024hallusionbench} systematically demonstrates that VLMs are prone to hallucinations when visual and textual cues conflict, a critical concern for engineering compliance applications.

Overall, existing literature suggests that while VLMs provide a strong foundation for technical and diagram understanding, their effectiveness is constrained by text-first preprocessing and limited spatial grounding. Most applications discussed above  remain focused on single-image interpretation and do not support scalable retrieval, evidence grounding, or compliance-oriented reasoning across large engineering drawing repositories. This motivates integrating vision-first retrieval with spatially grounded multimodal reasoning.

\subsection{Vision-First Document Retrieval and Multimodal RAG}\label{sec:lit-vfrag}

To overcome geometric loss, vision-first document retrieval methods index content directly within the visual embedding space. Models like ColPali \citep{faysse2025colpali} leverage multi-vector representations and late-interaction paradigms---initially developed by ColBERT \citep{khattab2020colbert}---to preserve spatial alignment globally without intermediate text extraction. These vision-first indexers confirm that layout-aware representations significantly improve retrieval on dense, visually complex documents \citep{kim2022donut}.

While early Multimodal RAG pipelines \citep{chen2022murag} still treated images merely as auxiliary grounding for textual retrieval, modern systems are beginning to capture direct diagrammatic semantics. Nonetheless, existing studies focus primarily on single-image understanding and isolated QA tasks, failing to address the robust structural linking required for compliance-oriented reasoning over large engineering archives. Addressing these limitations, we introduce an integrated visual-first RAG framework \textbf{(PlanSightRAG)}. It pairs late-interaction visual indexing with agentic VLM reasoning to meet the rigorous demands of cross-sheet engineering plan inspection.

\section{Methodology}\label{sec:methodology}

\subsection{Visual-First Multimodal RAG Framework Overview}\label{sec:meth-overview}

\begin{figure*}[t]
  \centering
  \resizebox{0.97\textwidth}{!}{\input{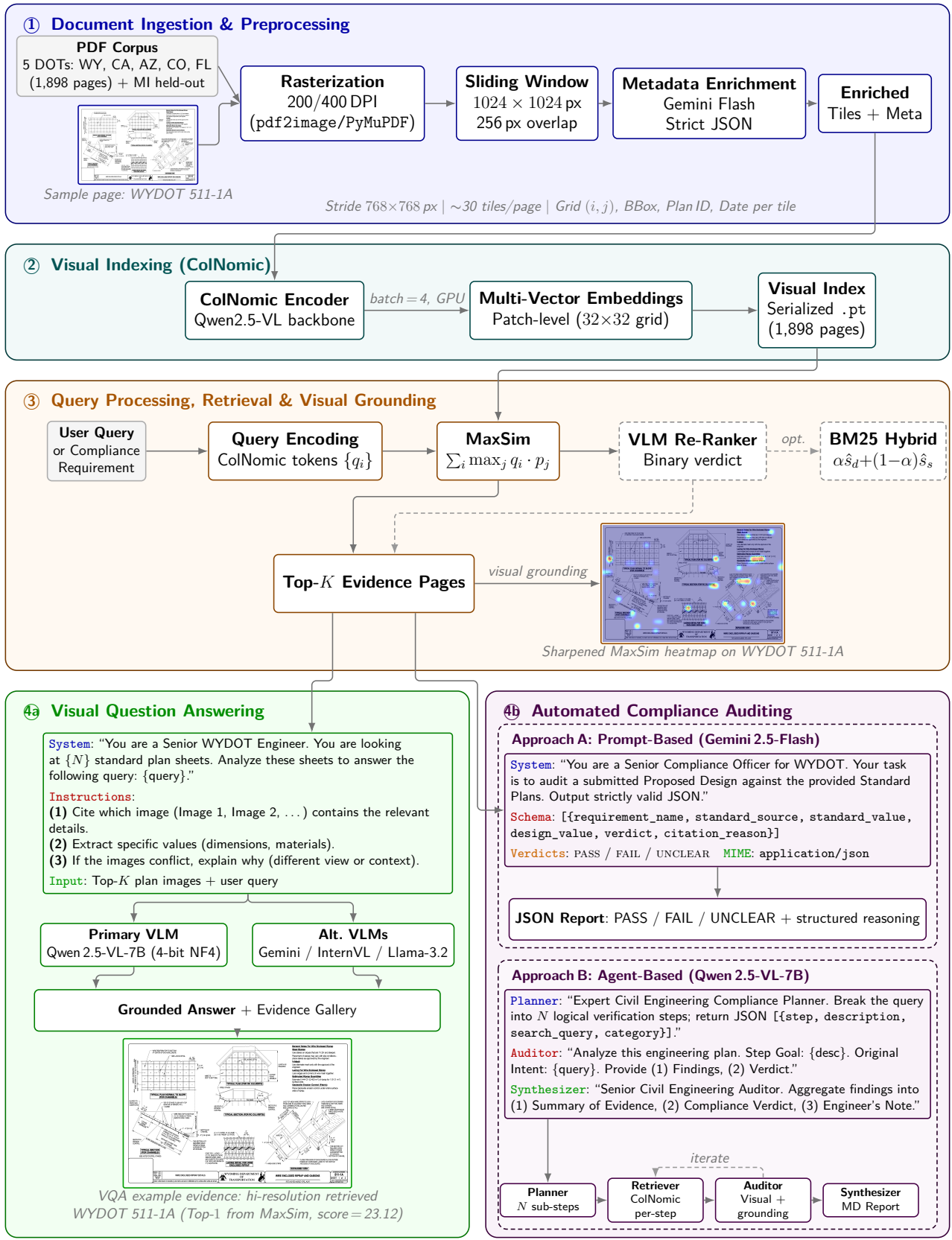}}
  \caption{Architecture of the Visual-First Multimodal RAG framework, organized into four phases: (1)~document ingestion with high-resolution tiling and metadata enrichment; (2)~ColNomic-3B visual indexing with multi-vector patch embeddings; (3)~MaxSim late-interaction retrieval and grounding (optional VLM re-ranking and BM25 fusion); and (4)~downstream tasks---(4a)~multi-VLM visual question answering with heatmap grounding and (4b)~agentic compliance auditing via the Planner--Retriever--Auditor--Synthesizer pipeline.}\label{fig:framework}
\end{figure*}

We propose a Visual-First Multimodal RAG architecture for question answering and compliance checking on civil engineering standard plans. As shown in Figure~\ref{fig:framework}, the framework is organized into four phases. \textbf{Phase~1} (Document Ingestion \& Preprocessing - section \ref{sec:meth-ingest}) rasterizes each PDF sheet for indexing. We use 200~DPI for the deployed full-page index used in all headline results, and 400~DPI for the optional tiling path evaluated as an ablation. When tiling is enabled, each sheet is split into overlapping $1024\times1024$ tiles with a 256-px overlap. Finally, the corpus is enriched with document-level metadata extracted by a vision--language model. \textbf{Phase~2} (Visual Indexing - section \ref{sec:meth-colpali}) encodes every tile with ColNomic-3B into multi-vector late-interaction patch embeddings at dynamic resolution and persists them as a reusable visual vector store. \textbf{Phase~3} (Query Processing, Retrieval \& Visual Grounding - section \ref{sec:meth-retrieval-grounding}) performs retrieval using MaxSim late-interaction scoring. It can optionally apply VLM cross-encoder re-ranking and BM25 hybrid fusion to refine the results. This phase also produces sharpened MaxSim heatmaps, which highlight the specific plan regions that drive each retrieval decision. \textbf{Phase~4} splits into two downstream tasks: Phase~4a (Visual Question Answering - section \ref{sec:meth-vqa}) generates grounded answers with a multi-VLM pool of Qwen~2.5-VL-7B, Qwen~2.5-VL-72B, and InternVL-2.5-8B over the Top-$K$ retrieved pages. Phase~4b (Automated Compliance Checking section - \ref{sec:meth-compliance}) runs a prompted Qwen~2.5-VL-72B auditor alongside an agentic Planner--Retriever--Auditor--Synthesizer pipeline to produce structured, evidence-grounded compliance reports. Qwen~2.5-VL-7B handles high-resolution VQA; Qwen~2.5-VL-72B handles long-context, multi-sheet compliance with enforced JSON output; and the remaining models are evaluated as alternative generators. The implementation details is discussed in section \ref{sec:meth-deploy} . A strict separation of concerns among retrieval, reasoning, grounding, and auditing reduces the risk of hallucinations and yields a transparent, evidence-grounded decision path.

\subsection{ Document Ingestion and Preprocessing}\label{sec:meth-ingest}

Standard Plans for Road and Bridge Construction from five US state DOTs(Wyoming, California, Arizona, Colorado and Florida) serve as the regulatory ground truth in our work. Each PDF document is rasterized page-by-page at 200~DPI, preserving line weights, hatch patterns, dimension chains, symbols, and annotations that are commonly lost in OCR. Pages are treated as independent visual units; OCR is intentionally excluded since the text-baseline results in Section~\ref{sec:res-comparison} (Table~\ref{tbl:comparison}) confirm label fragmentation in their geometric context(for example, the strongest modern text retriever BGE-M3 + OCR plateaus at 36.79\% Recall@5 against ColNomic's 92.69\% on the 424-pair test split). Preprocessing is limited to format conversion and resolution standardization (no cropping, segmentation, or manual annotation). Regarding resolution choices, we used \textbf{200~DPI full-page} for the visual index, all headline retrieval, and compliance-pipeline retrieval, and the \textbf{{400~DPI}} sliding-window tiling path is the only exception and is evaluated solely as an ablation (Section~\ref{sec:failure-analysis}). The combined index spans 1,898 pages: WYDOT (237), Caltrans~2025 (638), Arizona DOT~2025 (181), Colorado DOT~2025 (62), and Florida DOT~2026 (780), each deduplicated to a single publication year to remove near-identical revisions. A held-out Michigan DOT corpus (298 pages) is ingested in the same way and used exclusively for zero-shot cross-agency transfer evaluation.

\begin{figure*}[t]
  \centering
  \resizebox{\linewidth}{!}{\input{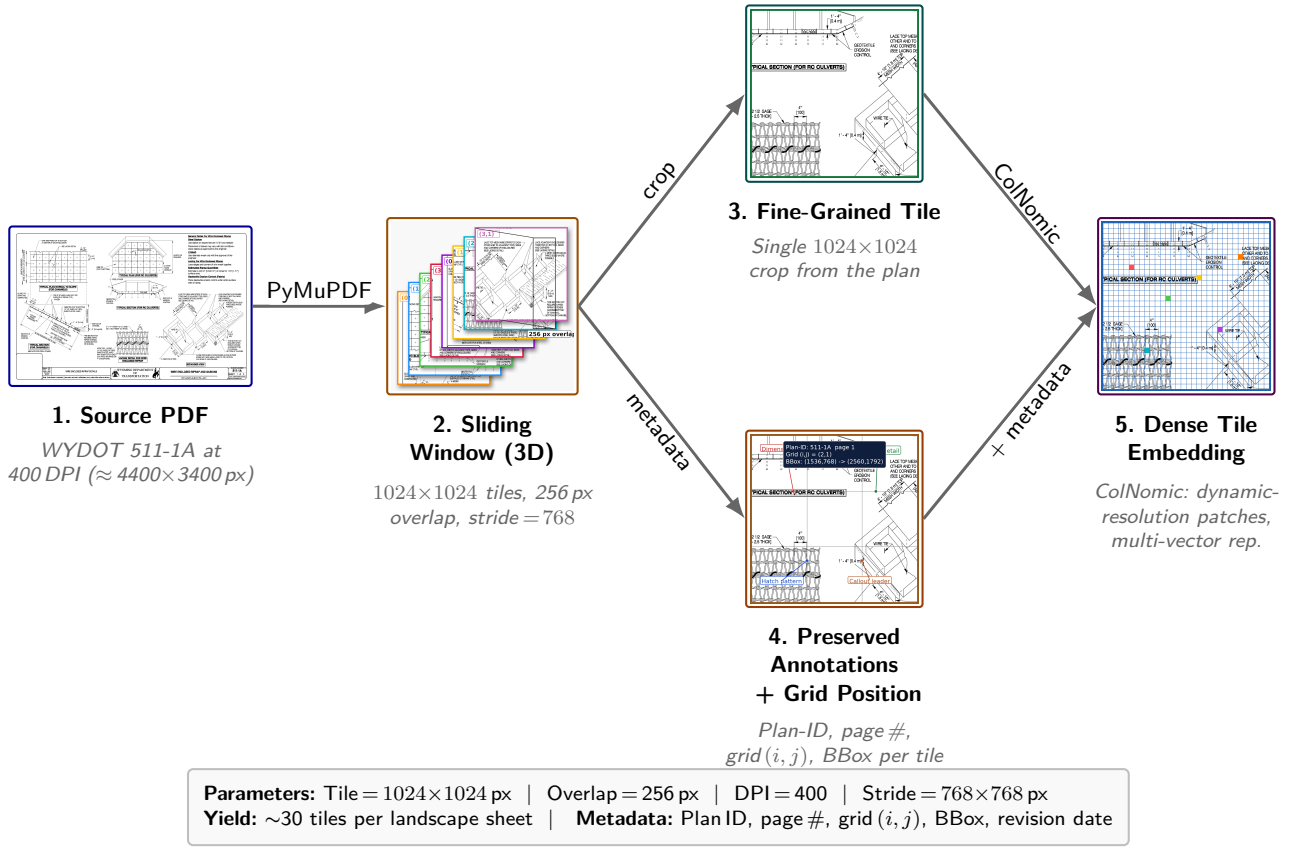}}
  \caption{Advanced Sliding Window Technique. Engineering sheets are rasterized at 400~DPI and decomposed into overlapping $1024 \times 1024$ tiles with 256-pixel overlap, preserving fine-grained annotations and enabling sub-page retrieval granularity.}\label{fig:tiling}
\end{figure*}

Even with a higher-resolution backbone, full-page indexing of very large multi-view sheets caps the effective per-region resolution, downsampling high-DPI engineering drawings until dimension text becomes illegible, thin lines vanish, and hatching merges. We therefore adopt a sliding-window tiling scheme (Figure~\ref{fig:tiling}) that preserves native-resolution detail by indexing at the tile level.

\paragraph{\textbf{Tiling Procedure:}} Each page is rasterized at 400 DPI with \texttt{PyMuPDF} and decomposed into overlapping tiles with a size $1024\times1024$, an overlap of 256 px (25\%), and a stride of $768\times768$, yielding $\lceil (W-256)/768 \rceil \times \lceil (H-256)/768 \rceil$ tiles, where $W,H$ are the full-resolution dimensions. A typical landscape sheet ($\approx$4400$\times$3400) produces $\approx$ 30 tiles.

\paragraph{\textbf{Metadata and Retrieval:}} Each tile inherits its document-level metadata (plan~ID, revision date, category, keywords) and adds tile-level fields: grid position $(i,j)$, bounding box $(x_{\text{left}}, y_{\text{top}}, x_{\text{right}}, y_{\text{bottom}})$, and a reference to the full-page image. Tiles are encoded with the same ColNomic retriever and scored individually, enabling sub-page retrieval granularity and precise spatial back-mapping for grounding.

\subsection{Visual Indexing}\label{sec:meth-colpali}

Each page (and each tile) is encoded with a pretrained ColNomic-3B retriever \citep{nomic2025colnomic} (\texttt{nomic-ai/\allowbreak colnomic-\allowbreak embed-\allowbreak multimodal-3b}), a multi-vector late-interaction model on the Qwen2.5-VL backbone \citep{bai2023qwenvl,vaswani2017attention}. Rather than compressing a page into a single vector, it produces a multi-vector embedding over a dynamic-resolution patch grid. This approach preserves spatial locality, allowing tables, legends, callouts, and geometric details to remain separately addressable. Such granularity is critical for engineering drawings, where compliance-relevant information is spatially distributed and cannot be summarized holistically. Pages are uniformly encoded in batches of four during inference. Each index entry stores the CPU-resident embedding tensor alongside a metadata record extracted once per PDF by a vision-language model under strict JSON output (with a filename-parsing fallback)—document-level fields (plan ID, revision date, title, category, keywords, source agency) and page-level fields (image path, page number, unique page ID). All 1,898 in-distribution pages are serialized as a single joint index, while the held-out Michigan DOT corpus (298 pages) is encoded identically into a separate index used only at zero-shot evaluation. The joint index supports cross-agency retrieval without per-agency partitioning, and decoupling indexing from downstream reasoning lets the same embeddings be reused across retrieval, VQA, and compliance without re-encoding.

\subsection{Query Processing, Retrieval \& Visual Grounding}\label{sec:meth-retrieval-grounding}

At query time, the ColNomic retriever embeds the query into token-level vectors $Q=\{q_1,\ldots,q_m\}$; each indexed page is represented by patch vectors $P=\{p_1,\ldots,p_n\}$. Retrieval uses a ColBERT-style \citep{khattab2020colbert,santhanam2022colbertv2} MaxSim late-interaction score computed in three steps:
\begingroup
\setlength{\jot}{3pt}
\begin{gather}
S = Q \cdot P^{\top} \label{eq:interaction}\\
\text{MaxSim}(q_i, P) = \max_{j=1,\ldots,n} \; q_i \cdot p_j \label{eq:maxsim}\\
\text{Score}(Q, P) = \sum_{i=1}^{m} \text{MaxSim}(q_i, P) \label{eq:score}
\end{gather}
\endgroup

Each query token independently aligns with its most relevant visual patch, which is essential when a dimension value, a legend symbol, and a table entry must all contribute to a page's relevance—cues that dense single-vector retrieval would average out. Pages are ranked by score, and the Top-$K$ are returned as visual evidence.

To optimize raw retrieval performance and adapt to mixed visual-lexical queries, we evaluate two enhancements layered on top of the base ColNomic retriever without modifying the visual index.

\paragraph{\textbf{VLM Cross-Encoder Re-Ranking:}} ColNomic returns Top-$K$ candidates (Stage~1). VLM then re-ranks each candidate with a binary prompt---\emph{``Does this engineering plan contain the specific data to answer the query?''}---retaining only positively verified candidates (Stage~2). Images are processed sequentially to bound GPU memory, with early stopping after two positive verifications. This yields a more discriminative relevance signal than bi-encoder similarity at the cost of bounded additional latency ($\le K$ passes, $\approx$ $0.3$~s each; Appendix~\ref{app:latency}).

\paragraph{\textbf{Hybrid Dense--Sparse Fusion:}} To complement visual similarity with lexical precision on alphanumeric queries (plan IDs, note references), ColNomic dense scores are fused with BM25 sparse scores over the extracted metadata (plan~ID, title, category, filename). Both vectors are min-max normalized to $[0,1]$ and combined:
\begin{equation}\label{eq:hybrid}
\text{Score}_{\text{hybrid}} = \alpha \cdot \hat{s}_{\text{dense}} + (1 - \alpha) \cdot \hat{s}_{\text{sparse}}
\end{equation}
with $\alpha \in [0,1]$ controlling the visual/lexical balance.

Beyond raw retrieval performance, ColNomic's patch-level late interaction reveals which regions drove each retrieval decision, making the architecture inherently explainable (unlike opaque text-in/text-out RAG).

\paragraph{\textbf{Sharpened MaxSim Heatmap:}} For each query page pair, patch-level MaxSim scores are reshaped into a 2D map aligned with the page layout and sharpened in two stages. In the first stage, only the top 5\% of patches are retained (below-95th-percentile scores are clamped to the minimum to suppress diffuse activations), and in the second stage, normalized values are raised to $\gamma = 3.0$ to concentrate emphasis on the strongest regions. The result is bicubic-upscaled to the original resolution, colorized with JET, and blended at $\alpha = 0.4$ over the page.

\paragraph{\textbf{Region-Level Grounding:}} High-activation patches are additionally scored against OCR-detected text regions. The Top-$K$ with the highest mean activation is returned as bounding-box evidence, pinpointing specific annotations, dimension labels, notes, or graphical elements behind the retrieval decision. The same mechanism attaches to every VQA answer and to each auditor step in the agentic pipeline, yielding a persistent visual evidence trail.

\subsection{Visual Question Answering (VQA) Pipeline}\label{sec:meth-vqa}

Given a query, the Top-$K$ plan pages returned by Section~\ref{sec:meth-retrieval-grounding} are passed to a vision--language model (Qwen~2.5-VL \citep{bai2023qwenvl}). The multimodal prompt consists of (i)~a strict system role (``You are an expert Civil Engineer for WYDOT; answer only from the provided Standard Plan images''), (ii)~the Top-$K$ retrieved images as explicit visual inputs, (iii)~lightweight text anchors (plan ID, reference index), and (iv)~the user query:
\begin{equation}\label{eq:generation}
\hat{y} = \text{VLM}\bigl(x, \{I_1, I_2, \ldots, I_K\}, I_{\text{user}}\bigr)
\end{equation}
where $x$ is the query, $\{I_k\}$ are the retrieved images, and $I_{\text{user}}$ is an optional user-supplied diagram. Decoding uses a fixed token budget; the output pairs a natural-language answer with an evidence gallery of supporting plan sheets, allowing users to verify answers directly against the source visuals.

\subsection{Automated Compliance Checking Workflow}\label{sec:meth-compliance}

For design-vs-standard auditing, a dedicated Qwen 2.5-VL-72B agent is configured as a WYDOT auditor. Its multimodal prompt includes (i)~Top-$K$ retrieved standard-plan images as regulatory evidence, (ii)~the proposed design image, (iii)~audit rules on dimensional consistency, unit correctness, labeling, and geometric logic, and (iv)~an enforced JSON output schema. Each audit item is classified under \textsc{unit}, \textsc{dimension}, \textsc{label}, or \textsc{logic}, reporting the applicable requirement, the observed design condition, and a verdict in \{\textsc{pass}, \textsc{fail}, \textsc{unclear}, \textsc{hallucination\_suspected}\}. This yields a structured, repeatable report that can be directly exported to downstream review tools.

While this configuration is effective for localized checks, single-pass RAG cannot handle compliance queries that span multiple plans and regulatory rules. We therefore introduce an \emph{Agentic Compliance} pipeline (Figure~\ref{fig:agentic}) with four specialized agents. While the framework is model-agnostic and supports lightweight backbones (such as Qwen~2.5-VL-7B for resource-constrained local deployment), we scale the backbone to Qwen~2.5-VL-72B for the primary high-accuracy evaluations reported in Section~\ref{sec:res-compliance-cad}:

\begin{enumerate}
    \item \textbf{Planner:} Decomposes a high-level query (e.g., ``Is the geotextile erosion control placement for RC Culverts consistent with fill slope requirements?'') into $N$ structured JSON steps, each specifying a focused search query, the expected plan category, and the information to verify.
    \item \textbf{Retriever:} Executes MaxSim retrieval per step, returning the most relevant plan page and its metadata.
    \item \textbf{Auditor:} Takes the step description, retrieves the image, and the original query; performs visual analysis; and emits a structured finding with a preliminary verdict.
    \item \textbf{Synthesizer:} Aggregates step findings into a consolidated report with a final verdict, a per-step evidence summary, and an engineer's note flagging ambiguities.
\end{enumerate}

Every audit step emits a sharpened MaxSim heatmap overlaid on the retrieved page. The pipeline outputs a machine-readable Markdown report interleaving findings, verdicts, and grounding images (plan~ID, auditor findings, evidence paths) for direct ingestion by engineering review workflows.

\begin{strip}
  \vspace{1em} 
  \centering
  \resizebox{\linewidth}{!}{\begin{tikzpicture}[
  >=Latex, font=\sffamily, thick,
  module/.style={draw=green!50!black, fill=green!5, rounded corners,
                 minimum width=3.3cm, minimum height=1.5cm, align=center,
                 inner sep=1ex, font=\sffamily\normalsize},
  data/.style={draw=blue!50!black, fill=blue!5, rounded corners,
               minimum width=2.2cm, minimum height=1.5cm, align=center,
               font=\sffamily\normalsize},
  prompt/.style={draw=black!40, fill=yellow!10, rounded corners=0.5ex,
                 align=left, inner sep=4pt, font=\sffamily\scriptsize,
                 text width=4.4cm}, 
  imgnode/.style={inner sep=1.5pt, draw=#1!60!black, thick,
                  rounded corners=0.3ex},
  arrow/.style={->, very thick, draw=black!60, rounded corners=3pt},
  arrowlbl/.style={font=\sffamily\footnotesize, fill=white, inner sep=1pt, align=center},
  annot/.style={font=\sffamily\footnotesize\itshape, text=black!50,
                align=center, text width=4.8cm},
]

\node[draw=purple!60, fill=purple!5, rounded corners, align=center,
      inner sep=6pt, text width=21.0cm, font=\sffamily\normalsize,
      anchor=north] (qbanner) at (10.1, 7.0)
  {\textbf{Sample Compliance Query:} \textit{``Does the proposed horizontal wire stiffener connection detail (18-gauge ties at 2-ft spacing) conform to WYDOT gabion wall standards, and is the 3:1 back-slope drainage acceptable?''}};

\node[data]   (query)  at (-1.0, 3.4) {\textbf{User}\\\textbf{Query}};

\node[module, fill=orange!10, draw=orange!70!black]
              (planner) at (3.0, 3.4)
              {\textbf{Planner}\\Qwen\,2.5-VL-72B\\decompose $\to N$ steps};

\node[module] (retriever) at (7.8, 3.4)
              {\textbf{Retriever}\\ColNomic MaxSim\\per-step search};

\node[module] (auditor) at (12.8, 3.4)
              {\textbf{Auditor}\\Qwen\,2.5-VL-72B\\visual analysis};

\node[module, fill=green!15, draw=green!60!black]
              (synth) at (17.6, 3.4)
              {\textbf{Synthesizer}\\Qwen\,2.5-VL-72B\\aggregate findings};

\node[data]   (report) at (21.2, 3.4)
              {\textbf{Audit}\\\textbf{Report}};

\draw[draw=orange!80!black, dashed, thick, rounded corners=2ex,
      fill=orange!5, fill opacity=0.25]
  (5.9, 5.1) rectangle (14.7, 2.2);
\node[font=\sffamily\normalsize\bfseries, text=orange!80!black,
      anchor=north east] at (14.6, 5.1)
  {Loop: Step $i = 1 \dots N$};

\draw[arrow] (query.east)     -- (planner.west);
\draw[arrow] (planner.east)   -- (retriever.west);

\draw[arrow] (retriever.east) -- node[arrowlbl, above]
             {visual\\[-0.5ex]evidence} (auditor.west);

\draw[arrow] (auditor.east)   -- (synth.west);
\draw[arrow] (synth.east)     -- (report.west);

\draw[arrow, draw=orange!70!black, densely dashed]
  (auditor.north) -- ++(0, 0.9) -| (retriever.north)
  node[arrowlbl, pos=0.25, above, text=orange!70!black] {next step};

\node[prompt, anchor=north, text width=7.5cm] (pp) at (1.5, 1.9)
  {\textcolor{blue!70}{\texttt{Prompt:}} ``You are an expert Civil Engineering Compliance Planner. Given the high-level compliance query, break it down into $N$ logical verification steps. Each step must identify (1) what information to look for, (2) which standard-plan category it belongs to. Return \texttt{JSON} only.''\\[2pt]
   \textcolor{red!70}{\texttt{Sample Output:}} \texttt{[\{step:1, desc:``wire stiffener detail'', search:``gabion wire tie'', category:``Retaining Wall''\}, \{step:2, desc:``back-slope drainage'', search:``3:1 slope drain'', category:``Drainage''\}]}};

\node[prompt, anchor=north] (rp) at (7.8, 1.9)
  {\textcolor{blue!70}{\texttt{Search (Step\,1):}} ``gabion wire tie''\\
   \textcolor{red!70}{\texttt{Top-1:}} WYDOT 511-1A p.\,1\\
   \phantom{\texttt{Top-1: }}MaxSim score = 23.12\\[3pt]
   \textcolor{blue!70}{\texttt{Search (Step\,2):}} ``3:1 slope drain''\\
   \textcolor{red!70}{\texttt{Top-1:}} WYDOT 203-2A p.\,1\\
   \phantom{\texttt{Top-1: }}MaxSim score = 19.87};

\node[prompt, anchor=north] (ap) at (12.8, 1.9)
  {\textcolor{blue!70}{\texttt{Prompt:}} ``Analyze this engineering plan. Step Goal: \{description\}. Original Intent: \{query\}. Provide (1)~\emph{Findings}: specific values/rules; (2)~\emph{Verdict}: does it comply or provide enough info?''\\[2pt]
   \textcolor{red!70}{\texttt{Step\,1:}} ``\#9 wire ties @ 2-ft o.c.\ per 511-1A~$\Rightarrow$ \textbf{Complies}.''\\
   \textcolor{red!70}{\texttt{Step\,2:}} ``3:1 slope OK; toe drain required per 203-2A.''};

\node[prompt, anchor=north] (sp) at (17.6, 1.9)
  {\textcolor{blue!70}{\texttt{Prompt:}} ``Senior Civil Engineering Auditor. Given per-step findings, produce (1)~\emph{Summary of Evidence}, (2)~\emph{Compliance Verdict}, (3)~\emph{Engineer's Note}.''\\[2pt]
   \textcolor{red!70}{\texttt{Output:}} Structured Markdown report with per-plan citations, overall verdict, and actionable engineer's note.};

\draw[densely dotted, thick, draw=black!40] (planner.south)   -- (planner.south |- pp.north);
\draw[densely dotted, thick, draw=black!40] (retriever.south) -- (rp.north);
\draw[densely dotted, thick, draw=black!40] (auditor.south)   -- (ap.north);
\draw[densely dotted, thick, draw=black!40] (synth.south)     -- (sp.north);

\node[module, fill=cyan!8, draw=cyan!50!black,
      minimum height=1.1cm, minimum width=5.0cm]
     (ground) at (10.3, -3.2)
     {\textbf{Grounding Module}\\Sharpened MaxSim $\to$ Heatmap Overlay};

\draw[arrow, draw=cyan!60!black] (rp.south) -- ++(0,-0.3) -| ([xshift=-1.5cm]ground.north);
\draw[arrow, draw=cyan!60!black] (ap.south) -- ++(0,-0.3) -| ([xshift=1.5cm]ground.north);

\node[imgnode=orange] (ev1) at (6.8, -6.8)
  {\includegraphics[width=5.2cm, height=3.0cm, keepaspectratio]{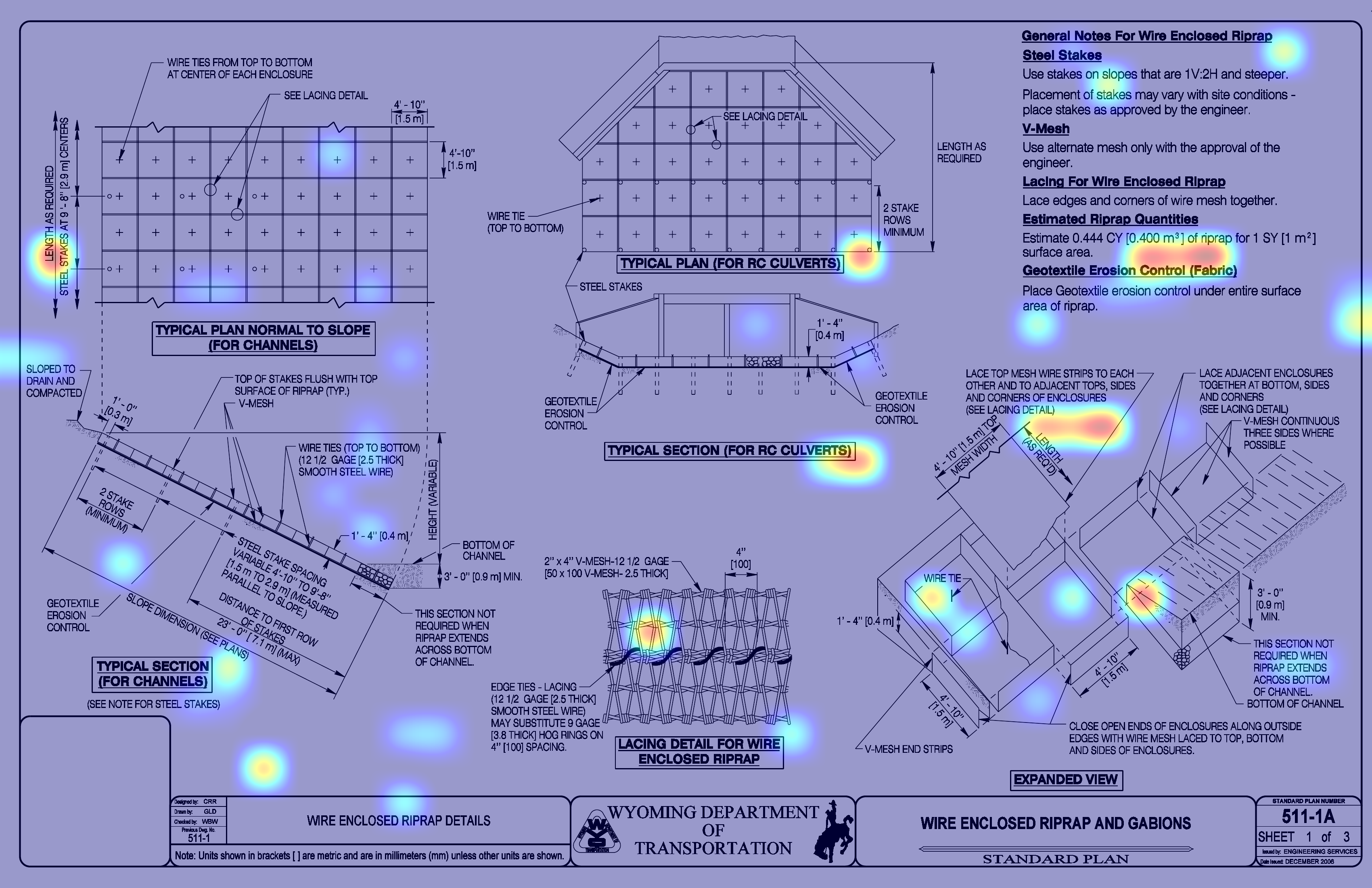}};
\node[annot, below=0.05cm of ev1] (ev1lbl)
  {Step\,1 evidence: WYDOT 511-1A\\(gabion wire-stiffener detail)};

\node[imgnode=orange] (ev2) at (13.8, -6.8)
  {\includegraphics[width=5.2cm, height=3.0cm, keepaspectratio]{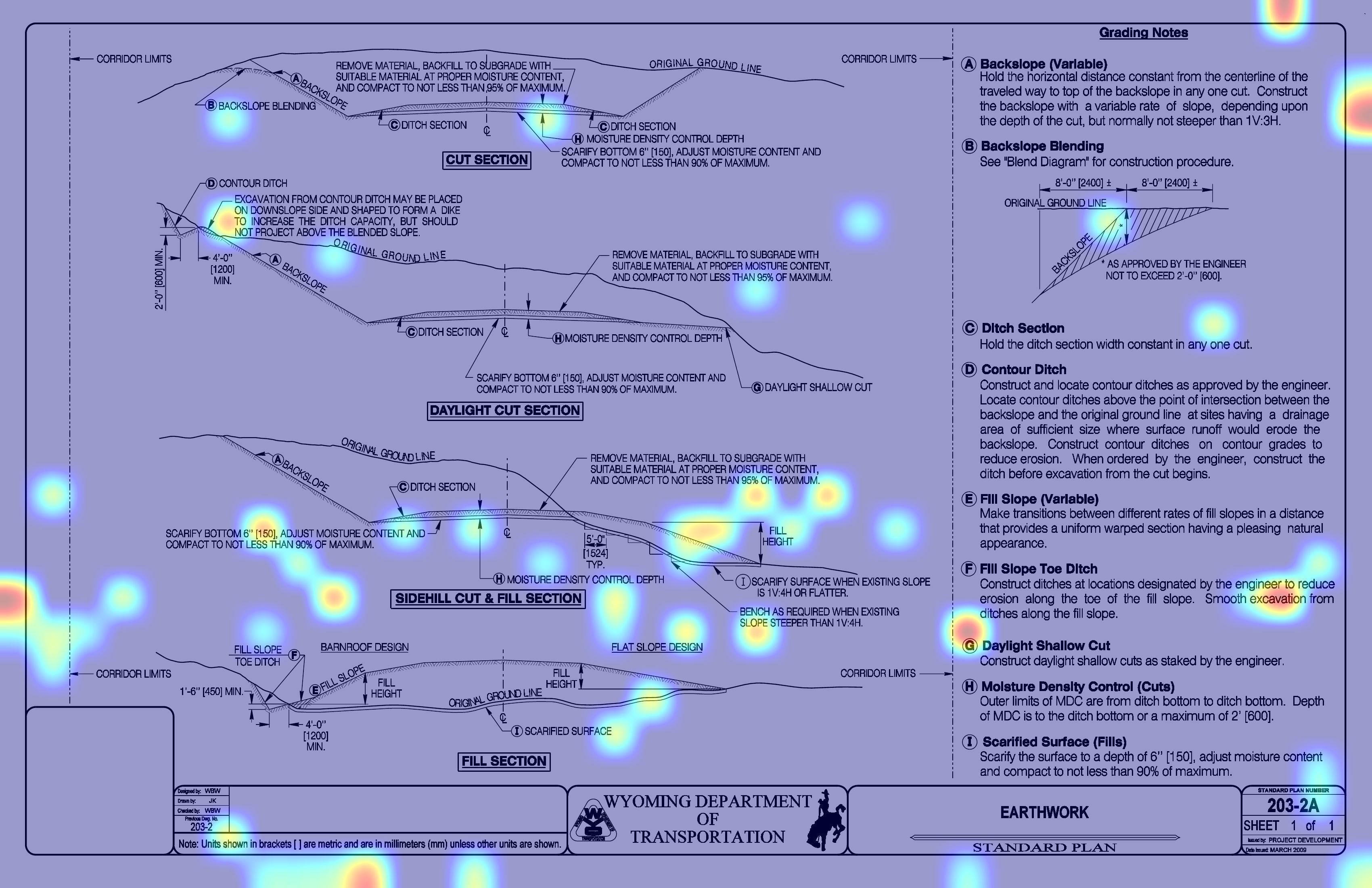}};
\node[annot, below=0.05cm of ev2] (ev2lbl)
  {Step\,2 evidence: WYDOT 203-2A\\(back-slope / drainage)};

\draw[arrow, draw=orange!70!black]
  ([xshift=-1.0cm]ground.south) -- ++(0,-0.4) -| (ev1.north);
\draw[arrow, draw=orange!70!black]
  ([xshift=1.0cm]ground.south) -- ++(0,-0.4) -| (ev2.north);

\node[imgnode=blue] (rpt) at (21.2, -6.4)
  {\includegraphics[width=3.8cm, height=4.5cm, keepaspectratio]{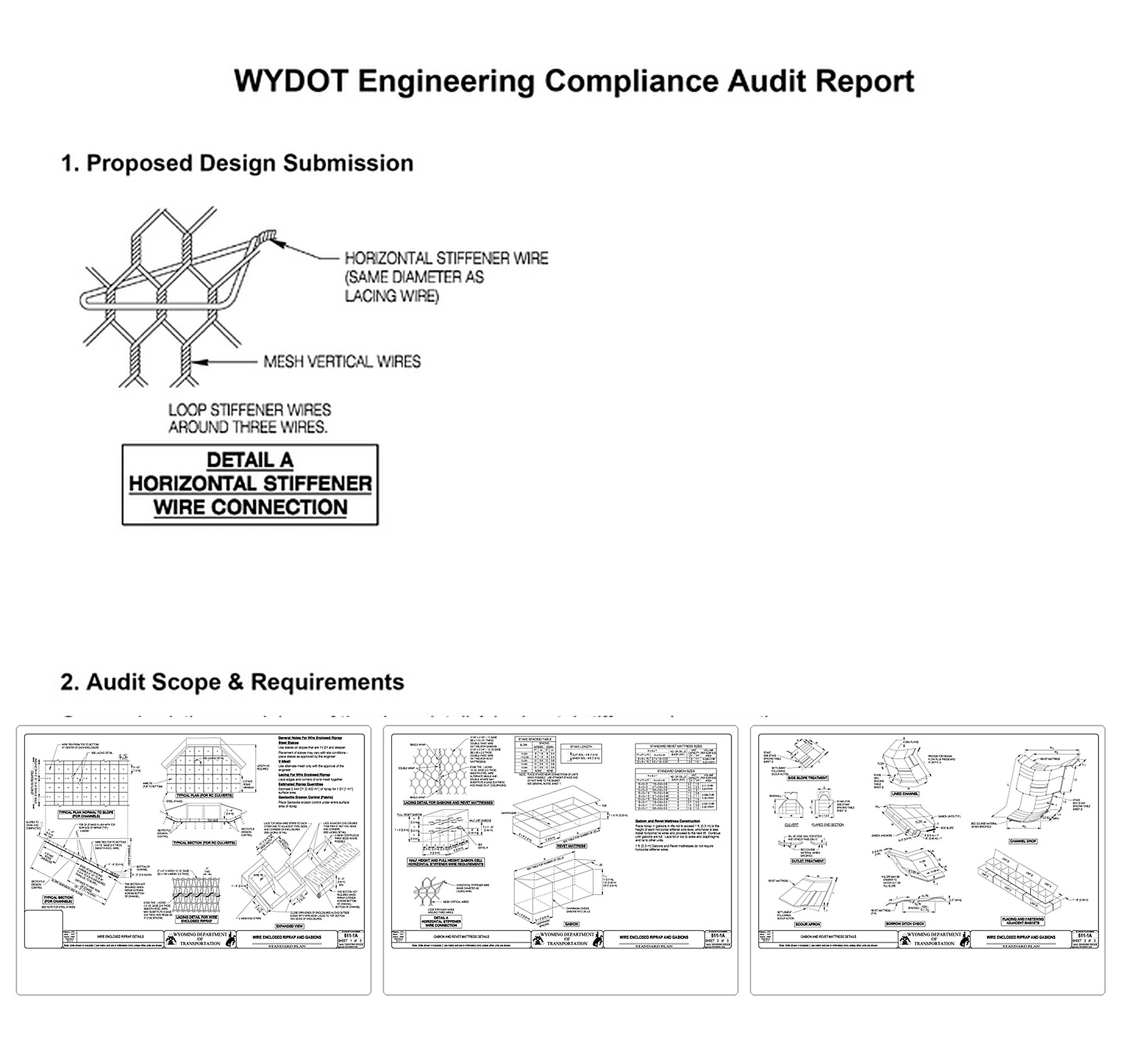}};
\node[annot, below=0.05cm of rpt, text width=3.8cm]
  {Generated compliance audit report (Markdown + citations)};

\draw[arrow, draw=blue!50!black]
  (report.south) -- (rpt.north);

\end{tikzpicture}}
  \captionof{figure}{Agentic Compliance Framework: Planner--Retriever--Auditor--Synthesizer pipeline for multi-step, evidence-grounded compliance verification.}\label{fig:agentic}
  \vspace{1em} 
\end{strip}

\subsection{Implementation and Deployment Details}\label{sec:meth-deploy}

The system runs on a single GPU to match practical agency deployments. Indexing and retrieval are executed offline; VQA and compliance inference run at query time. Qwen~2.5-VL uses 4-bit BitsAndBytes quantization (NF4 weights, bfloat16 compute), which reduces memory usage and allows several full-resolution plan images to fit in a single prompt. Models are loaded in inference mode only. The resulting efficiency–accuracy trade-off is managed through strict, evidence-grounded prompting, bounded generation lengths, and a retrieval-first design, ensuring that answers are anchored in retrieved evidence rather than priors.

\paragraph{\textbf{Deployment latency:}} On the single-GPU target, indexing the $1,898$-page corpus is a one-time offline cost ($7.3$~min, $4.34$~pages/s); at query time, ColNomic retrieval over the full index is $\approx$$0.10$~s (p50). The optional cross-encoder re-ranker adds $\approx$$0.3$~s per candidate, single-shot VQA is $2$--$5$~s, and the agentic compliance pipeline averages $60.9$~s/query because the Planner expands each verdict into $\sim$$8.6$ sequential VLM steps—the dominant cost and the target of the step-pruning optimization in Section~\ref{sec:future}. Full timings are in Appendix~\ref{app:latency} (Tables~\ref{tbl:efficiency},~\ref{tbl:latency}).

\section{Evaluation Setup}\label{sec:eval-setup}

\subsection{Benchmark Dataset Construction}\label{sec:eval-dataset}

To rigorously evaluate the framework, we construct (i)~a 4,056-pair benchmark generated via an iterative refine-loop QnA procedure across five state DOTs, (ii)~a held-out Michigan DOT zero-shot transfer set, and (iii)~a CAD-generated compliance test set used to validate the agentic verdict pipeline against known ground truth.

\paragraph{\textbf{Five-DOT Visual Index (1,898 Pages):}} The retrieval index aggregates standard plan PDFs from five US state DOTs, deduplicated to a single year per agency to remove near-identical revisions: WYDOT (237~pages), Caltrans~2025 (638~pages), Arizona DOT 2025 (181~pages), Colorado DOT~2025 (62~pages), and Florida DOT~2026 (780~pages), for a combined 1,898 pages. Each page is rendered at 200~DPI and stored as a multi-vector record with metadata (agency, plan ID, sheet title) extracted by a Qwen2.5-VL-7B metadata extractor. During benchmark \emph{construction} (the refine loop below), the curation index is built with ColPali~\citep{faysse2025colpali}; all \emph{reported} retrieval results re-embed the identical pages with the adopted ColNomic-3B backbone (Section~\ref{sec:res-comparison}).

\paragraph{\textbf{Refine-Loop QnA Generation:}} Manually authoring large QnA benchmarks for engineering plans is cost-prohibitive—prior engineering plan QA sets are typically limited to tens of manually verified pairs. We replace manual authoring with an iterative procedure: for each target page and each of four reasoning categories (Dimensional Accuracy, Visual Interpretation, Logical Reasoning, Hallucination Rate), a drafter VLM (Qwen2.5-VL-72B or Llama-3.2-90B-Vision in 4-bit NF4) writes a question conditioned on the page image, ColPali retrieves over the 1,898-page index, and if the target page is outside top-5 the drafter rephrases with a page-specific anchor (a plan ID, a labeled value, or a quoted note phrase). Up to five rephrase attempts are allowed per question (eight for the small 62-page Colorado corpus, with a stricter two-anchor rephrase). A separate Qwen-VL-7B verifier subsequently filters out questions whose answers cannot be derived from the page (yes/no rubric). The procedure yields \textbf{4,056 verified question–answer pairs} that pass both the refine-loop and the independent verifier. A further 93 verified pairs are generated in the same way over the held-out Michigan DOT corpus for zero-shot transfer evaluation.

\paragraph{\textbf{Construct validity of the curation loop:}} Because the refine loop accepts or rephrases a question according to whether \emph{ColPali} ranks the target page in the top-5, the benchmark could, in principle, favor ColPali-retrievable phrasings and inflate any retriever evaluated on it. Two facts rule this out as a material confound. First, $23.9\%$ of the finalized pairs ($968/4,056$; $95/424$ on the page-disjoint test split) are ColPali \emph{misses}—questions ColPali never retrieved even after five rephrases—so the benchmark is not restricted to ColPali-friendly items. Second, on exactly these ColPali-miss test questions, the adopted ColNomic-3B retriever still attains \textbf{77.89\% Recall@5} (vs.\ $96.66\%$ on the ColPali-hit subset), recovering the large majority of ColPali's own failures. A benchmark that merely encoded ColPali's inductive biases could not be solved by a \emph{different} retriever on ColPali's failure set; the adopted retriever's advantage is therefore genuine rather than an artifact of the curation retriever.

\paragraph{\textbf{Manually-curated anchor set (78 WYDOT pairs):}} Prior to the machine-generated benchmark, we hand-authored and manually verified a set of \textbf{78 WYDOT QnA pairs} (29 Dimensional, 16 Visual, 17 Logical, 16 Hallucination) against the source standard plans. Because these pairs carry \emph{no} machine-generation provenance, they provide an independent check that the framework's headline behavior is not an artifact of the generate-then-verify pipeline. On this human-curated set—retrieving over the 237-page WYDOT-only slice of the index—the system attains \textbf{91.03\% Recall@5} (95\% bootstrap CI $[84.6, 96.2]$) and, with the same local Qwen2.5-VL-7B answerer and Qwen2.5-VL-72B judge used throughout, \textbf{80.77\%} end-to-end judge accuracy (CI $[71.8, 89.7]$); the per-category breakdown is reported in Table~\ref{tbl:legacy78q}. The retrieval profile matches the large machine-generated benchmark (Dimensional strongest, Hallucination/Visual hardest), confirming that the automatically-generated pairs reproduce the difficulty structure of the human-authored ones.

\begin{table}[!t]
\caption{Human-curated 78-pair WYDOT anchor set: per-category Recall@5 and end-to-end judge accuracy with $95\%$ bootstrap CIs ($10,000$ resamples, seed~42). Retrieval is over the 237-page WYDOT-only index; judge accuracy uses the local Qwen2.5-VL-7B answerer with a Qwen2.5-VL-72B judge (no proprietary API).}\label{tbl:legacy78q}
\centering\footnotesize
\setlength{\tabcolsep}{4pt}
\begin{tabular*}{\columnwidth}{@{\extracolsep{\fill}}lccc@{}}
\toprule
Category & $N$ & Recall@5 (\%) & Judge (\%) \\
\midrule
Dimensional   & 29 & 100.00 [100.0, 100.0] & 79.31 [62.1, 93.1] \\
Visual        & 16 & 87.50 [68.8, 100.0]  & 68.75 [43.8, 87.5] \\
Logical       & 17 & 94.12 [82.4, 100.0]  & 94.12 [82.4, 100.0] \\
Hallucination & 16 & 75.00 [50.0, 93.8]   & 81.25 [62.5, 100.0] \\
\midrule
\textbf{Overall}      & \textbf{78} & \textbf{91.03 [84.6, 96.2]} & \textbf{80.77 [71.8, 89.7]} \\
\bottomrule
\end{tabular*}
\end{table}

The four reasoning categories are defined as follows:
\begin{itemize}
    \item \textbf{Dimensional Accuracy:} Queries requiring precise extraction of numeric dimensions, spacing, depths, gauges, and material quantities from engineering annotations.
    \item \textbf{Visual Interpretation:} Queries testing recognition of non-textual semantics, including geometric configurations, layout-dependent meaning, symbol conventions, and spatial relationships.
    \item \textbf{Logical Reasoning:} Queries requiring multi-step deduction based on retrieved visual evidence, such as reconciling notes, cross-referencing views, or applying conditional specifications.
    \item \textbf{Hallucination Rate:} Adversarial queries where the correct answer is ``not specified,'' testing whether models avoid asserting unsupported facts when evidence is missing or ambiguous.
\end{itemize}

\paragraph{\textbf{Hallucination-Rate category:}} The verifier accepts a pair when the page genuinely supports the answer, which by construction retains predominantly \emph{answerable} lookups in this category; only a minority of finalized pairs carry a literal ``not specified'' gold answer. For \emph{retrieval} evaluation, this is immaterial---every pair has a well-defined gold page---so the category should be read as a spec/material-lookup stress set rather than a pure abstention test. Abstention behavior is instead measured directly on the CAD false-negative study (Section~\ref{sec:res-agentic}). The per-agency and per-category distribution of the 4,056-pair benchmark is reported in Table~\ref{tbl:cat-counts}.

\begin{table*}[!t]
\caption{Distribution of the 4,056-pair five-DOT benchmark by agency and reasoning category. The held-out Michigan DOT zero-shot set (93 pairs) is not included in this table.}\label{tbl:cat-counts}
\centering
\small
\begin{tabular*}{\textwidth}{@{\extracolsep{\fill}}lcccccc@{}}
\toprule
Category & WYDOT & Caltrans & AZDOT & CDOT & FDOT & Total \\
\midrule
Visual Interpretation & 201 & ~416 & 149 & ~85 & ~549 & 1,400 \\
Dimensional Accuracy  & 170 & ~287 & 144 & ~87 & ~327 & 1,015 \\
Logical Reasoning     & 172 & ~331 & 104 & 104 & ~446 & 1,157 \\
Hallucination Rate    & ~15 & ~192 & ~17 & ~~2 & ~258 & ~~484 \\
\midrule
\textbf{Total}        & \textbf{558} & \textbf{1,226} & \textbf{414} & \textbf{278} & \textbf{1,580} & \textbf{4,056} \\
\bottomrule
\end{tabular*}
\end{table*}

\paragraph{\textbf{CAD-Generated Compliance Test Set:}} To evaluate the agentic verdict pipeline against \emph{known} ground truth (so that judge errors can be unambiguously attributed to extraction, rule selection, or arithmetic), we construct a parameterized \emph{synthetic} drawing generator (\texttt{matplotlib}; schematic figures rather than CAD/DWG output) that produces engineering-plan drawings with title block, dimensioned section views, material callouts, and notes. The generator covers five archetypes (box culvert, guardrail post foundation, beam rebar detail, drainage inlet, sign post foundation) at three density tiers (single-view, multi-view + schedule table, and multi-component multi-plan), with controlled parameter values. For non-compliant instances, exactly one parameter violates a rule (e.g., cover $=1.0''<2.0''$~min, stirrup spacing $=18''>d/2{=}15''$~max). The full compliance ground-truth set comprises \textbf{(a)}~a 14-drawing pilot ($n{=}10$ single-doc spanning the 5 archetypes $+$ $n{=}4$ dense multi-view with the violation embedded in one schedule-table row), \textbf{(b)}~a \textbf{500-drawing} single-doc scale set (5 archetypes $\times$ 100 instances, 250 compliant $+$ 250 non-compliant, seven violation families), and \textbf{(c)}~a \textbf{100-drawing} multi-plan set requiring $N{\in}\{2,3,4,5\}$ separate standard plans for verdict, with \emph{no} plan IDs cited on the drawings (pure visual inference). Section~\ref{sec:res-compliance-cad} reports verdict accuracy on all three.

\paragraph{\textbf{Dataset Standardization:}} All tracks are unified into a single evaluation pipeline that standardizes input formats (question, ground truth, image file, category, and applicable rule thresholds), enabling consistent metric computation across all models and configurations.

\subsection{Evaluated Models}\label{sec:eval-models}

We evaluate vision–language models in two roles: as \emph{drafters} in the refine-loop QnA generation (Qwen2.5-VL-72B and Llama-3.2-90B-Vision) and as the \emph{compliance judge} in the agentic pipeline (Qwen2.5-VL-72B). The ColNomic visual retrieval backend is held fixed across all generator comparisons. Table~\ref{tbl:models} summarizes the models and their roles.

\begin{table}[!t]
\caption{Vision--language models used in the framework.}\label{tbl:models}
\centering
\resizebox{\linewidth}{!}{%
\begin{tabular}{@{}llcc@{}}
\toprule
Model & Type & Params & Tier \\
\midrule
Qwen2.5-VL-72B-Instruct        & Open-source   & 72B & Primary \\
Qwen2.5-VL-7B-Instruct         & Open-source   & 7B  & Primary \\
InternVL-2.5-8B                & Open-source   & 8B  & Primary \\
\bottomrule
\end{tabular}%
}
\end{table}

All open-source models are deployed with 4-bit NF4 quantization via BitsAndBytes to enable single-GPU inference. The evaluation pipeline is fully automated: for each model, the system loads the ColNomic retriever, performs visual retrieval for each benchmark question, generates an answer using the model under test, and computes evaluation metrics.

\subsection{Evaluation Metrics}\label{sec:eval-metrics}

Performance is reported using two complementary KPIs:

\begin{itemize}
    \item \textbf{Recall@5 (Retrieval Hit):} The percentage of queries for which the correct standard-plan page appears in the Top-5 retrieved results \citep{patel2022recall}. We report Recall@5 rather than Recall@K generically because the end-to-end pipeline always passes the Top-5 candidates to the generator. This measures the effectiveness of the visual retrieval layer independent of the generator.
    \item \textbf{Judge Accuracy:} An LLM-as-judge score \citep{zheng2023judging} where a local Qwen2.5-VL-72B judge compares the model's predicted answer against the ground truth and assigns a binary score (0 or 1). This measures end-to-end correctness, including both retrieval and reasoning.
\end{itemize}

\section{Results}\label{sec:results}

\subsection{Comparison with Existing Solutions}\label{sec:res-comparison}

To test H1, we pre-declare the acceptance threshold as a $\ge$15~pp Recall@5 margin over the strongest text-based and hybrid baselines and compare patch-level visual retrieval (ColPali) against a broad set of strong modern retrieval baselines, including the 2025--2026 ViDoRe state of the art: NVIDIA Nemotron-ColEmbed-8B/4B \citep{nvidia2026nemotroncolembed} (Qwen3-VL backbones) and ColNomic-3B/7B \citep{nomic2025colnomic}, along with ColQwen2.5-v0.2 and ColPali-v1.3 \citep{faysse2025colpali}, DSE-Qwen2-2B \citep{ma2024dse}, VisRAG-Ret \citep{yu2024visrag}, and BGE-M3 with OCR \citep{chen2024bgem3}—alongside seven legacy baselines (CLIP, LayoutLMv3, Nougat, Pix2Struct, UDOP, OCR+MiniLM, VisionRAG-Pyramid) and a binary-quantized HPC-ColPali variant. Every method is evaluated identically: the 424-pair page-disjoint test split is queried against the full 1,898-page five-DOT visual index. Table~\ref{tbl:comparison} reports Recall@5 for each method alongside index-size and retrieval-latency figures; per-agency and per-category bootstrap CIs are reported in Appendix~\ref{app:bench-stats}.

\paragraph{\textbf{On the choice of retriever:}} Several 2025--2026 retrievers outperform zero-shot ColPali on this benchma-rk----Nemotron-ColEmbed-8B reaches 95.28\% Recall@5 and ColNomic-3B 92.69\%, versus 76.89\% for ColPali. This reflects their newer, higher-resolution Qwen2.5/ Qwen3-VL backbones, whereas ColPali's PaliGemma encoder operates at a fixed $448{\times}448$ that coarsens the fine dimensions and notes on large standard-plan sheets.This is the same resolution limitation our high-resolution tiling strategy targets (Section~\ref{sec:failure-analysis}). The framework is deliberately retriever-agnostic: the indexing, tiling, visual grounding, and agentic compliance components are unchanged by the backbone, so any of these retrievers can drop in directly, and the per-agency gains transfer. Complementarily, a lightweight Apache-2.0 visual reranker (MonoQwen2-VL) applied to ColPali's top-20 lifts Recall@5 from 76.89\% to 85.14\% (Table~\ref{tbl:rerank})---recovering most of the gap to the stronger retrievers while keeping the stack fully open and deployable. We therefore \emph{adopt ColNomic-3B as the retrieval backbone}—the strongest openly licensed visual retriever on our benchmark and fully deployable—and report all main results with it; ColPali serves as a companion backbone for our binary-quantization study (HPC-ColPali, $16\times$ compression) and controlled LoRA analysis. The NVIDIA Nemotron-ColEmbed models reach even higher Recall@5 but are CC-BY-NC (research-only). Crucially, H1 concerns the gap to \emph{text-based and hybrid} pipelines, which every visual retriever here clears by a wide margin. For clarity on backbone attribution: all \emph{headline} retrieval and the rule-grounding and real-plan results use ColNomic-3B, whereas the re-ranking (Table~\ref{tbl:rerank}), high-resolution tiling (Section~\ref{sec:failure-analysis}), HPC binary-quantization, bootstrap-CI (Appendix~\ref{app:bench-stats}), and grounding-pilot (Appendix~\ref{app:grounding}) studies are companion analyses on the predecessor ColPali backbone, and each is labeled accordingly.

\begin{table*}[t]
\caption{Retrieval comparison on the 424-pair page-disjoint test split over the 1,898-page five-DOT index. Strong (modern) baselines on top, legacy/weak baselines below, and the predecessor ColPali backbone (zero-shot) at the bottom; the adopted retriever is ColNomic-3B (92.69\%, bolded). All numbers measured under an identical evaluation protocol; per-agency and per-category bootstrap CIs are reported in Appendix~\ref{app:bench-stats}.}\label{tbl:comparison}
\begin{tabular*}{\textwidth}{@{\extracolsep{\fill}}llllc@{}}
\toprule
Method & Modality & Mechanism & $N_{\text{idx}}$ & Recall@5 \\
\midrule
\multicolumn{5}{c}{\emph{Strong modern baselines (2024--2026)}} \\
Nemotron-ColEmbed-8B \citep{nvidia2026nemotroncolembed} & Visual (Patch) & Multi-vector late-int. & 1898 & 95.28\% \\
Nemotron-ColEmbed-4B \citep{nvidia2026nemotroncolembed} & Visual (Patch) & Multi-vector late-int. & 1898 & 92.92\% \\
ColNomic-7B \citep{nomic2025colnomic} & Visual (Patch) & Multi-vector late-int. & 1898 & 91.27\% \\
ColQwen2.5-v0.2 \citep{faysse2025colpali} & Visual (Patch) & Multi-vector MaxSim & 1898 & 87.26\% \\
DSE-Qwen2-2B \citep{ma2024dse} & Visual (Global) & Single-vector dense & 1898 & 22.17\% \\
VisRAG-Ret \citep{yu2024visrag} & Visual (Global) & Single-vector dense & 1898 & 53.77\% \\
BGE-M3 + OCR \citep{chen2024bgem3} & Text (OCR) & Dense Vector & 1898 & 36.79\% \\
{ColNomic-3B (Ours, adopted)} \citep{nomic2025colnomic} & Visual (Patch) & Multi-vector late-int. & 1898 & 92.69\% \\
\midrule
\multicolumn{5}{c}{\emph{Legacy baselines }} \\
CLIP ViT-B/32 \citep{radford2021learning} & Visual (Global) & Global Embedding & 1898 & 1.89\% \\
LayoutLMv3 \citep{huang2022layoutlmv3} & Visual (Layout) & Layout Embedding & 1898 & 0.00\% \\
Nougat-decode + MiniLM & OCR-free (Decode) & Decode + Dense & 1898 & 0.47\% \\
Pix2Struct-decode + MiniLM & OCR-free (Decode) & Decode + Dense & 1898 & 7.31\% \\
UDOP-decode + MiniLM & OCR-free (Decode) & Decode + Dense & 1898 & 0.00\% \\
OCR + MiniLM & Text (OCR) & Dense Vector & 1898 & 25.24\% \\
VisionRAG (Pyramid, RRF) & Hybrid (Text) & Structure-Aware RRF & 1898 & 23.11\% \\
\midrule
ColPali (predecessor backbone) \citep{faysse2025colpali} & Visual (Patch) & Multi-vector MaxSim & 1898 & 76.89\% \\
HPC-ColPali (Binary Quantized) & Visual (Patch) & Multi-vector MaxSim & 1898 & 65.09\% \\
\bottomrule
\end{tabular*}
\end{table*}

\begin{table}[!t]
\caption{Two-stage retrieve--rerank on the 424-pair test split. The predecessor ColPali backbone retrieves the top-20; MonoQwen2-VL-v0.1 (Apache-2.0 pointwise visual reranker) reorders them. The top-20 recall is the reranker's ceiling.}\label{tbl:rerank}
\centering
\small
\begin{tabular}{lc}
\toprule
Configuration & Recall@5 (\%) \\
\midrule
ColPali (Stage 1 only) & 76.89 \\
ColPali + MonoQwen2-VL rerank & 85.14 \\
\midrule
\emph{Top-20 ceiling} & \emph{86.79} \\
\bottomrule
\end{tabular}
\end{table}

\paragraph{\textbf{Global vs.\ Local Vision:}} The comparison between CLIP and ColPali reveals a stark ``granularity gap.'' CLIP, which compresses an entire document page into a single global vector, achieved Recall@5 of only 1.89\%, confirming that global embeddings fail to resolve the fine-grained details of engineering schematics. In contrast, patch-level ColPali achieves 76.89\% Recall@5, demonstrating that retaining spatial resolution is non-negotiable for this domain.

\paragraph{\textbf{Retrieval ceiling for text-centric baselines:}}
The strongest modern text retriever evaluated against the 1,898-page index is BGE-M3 + OCR at 36.79\% Recall@5, the upper bound of what an OCR-then-embed pipeline can deliver here, even after replacing legacy baselines with 2024–2025 state-of-the-art encoders. The hybrid VisionRAG pyramid, which explicitly models document structure, raises this only to 23.11\%. ColNomic's 92.69\% Recall@5 sits 55.90 pp above the strongest text retriever and 69.58 pp above the strongest hybrid, indicating that the information loss occurs at the OCR step itself rather than at the embedder.

\paragraph{\textbf{OCR-free document VLMs do not transfer:}} Three OCR-free document VLMs used as page-to-text encoders—Nougat \citep{blecher2023nougat}, Pix2Struct \citep{lee2023pix2struct}, and UDOP \citep{tang2023udop}, with text embedded by MiniLM, so only the page-to-text step varies—also collapse on engineering drawings (0.47\%, 7.31\%, and 0.00\% Recall@5). Pretrained on text-heavy corpora (papers, forms, screenshots), they emit empty or layout-less output on dense plan sheets, extending the granularity-gap argument to the broader OCR-free family.

\paragraph{\textbf{Efficiency vs.\ Accuracy:}} The binary-quantization variant (HPC-ColPali) compresses the visual index by $16.0\times$ (477.74~MB to 29.86~MB) and retains 65.09\% Recall@5—within 11.79~pp of full-precision ColPali.

\paragraph{\textbf{H1 verdict:}} H1 is \emph{supported}. The adopted ColNomic-3B retriever reaches 92.69\% Recall@5, which is 55.90~pp above the strongest text retriever (BGE-M3 + OCR, 36.79\%) and 69.58~pp above the strongest hybrid baseline (VisionRAG (Pyramid, RRF), 23.11\%). The pre-declared 15-pp threshold is cleared on both axes.

\subsection{Overall End-to-End Performance on the Five-DOT Benchmark}
\label{sec:res-overall}

Table~\ref{tbl:overall} presents the zero-shot retrieval performance of the adopted ColNomic-3B backbone on the full 4,056-pair five-DOT benchmark. This is the \emph{baseline} retrieval configuration; the controlled LoRA ablation in Section \ref{sec:res-retrieval-ablation} indicates that it is also the strongest deployment configuration among those we tested. End-to-end judge accuracy is measured separately on the CAD-generated compliance test set (Section~\ref{sec:res-compliance-cad}) because it has controlled ground-truth verdicts. Rank-sensitive metrics (Recall@1, MRR, nDCG@10) on the page-disjoint test split are reported in Appendix~\ref{app:extra} (Table~\ref{tbl:rankmetrics}). For clarity, three Recall@5 figures recur in this paper and each is reported against a distinct evaluation set: \textbf{91.47\%} on the full \emph{4,056-pair} benchmark (this section, our headline zero-shot number), \textbf{92.69\%} on the \emph{424-pair} page-disjoint test split (the controlled baseline comparison of Table~\ref{tbl:comparison} and the rank metrics), and \textbf{91.03\%} on the human-curated \emph{78-pair} WYDOT anchor set (Section~\ref{sec:eval-dataset}, Figure~\ref{fig:intro_comparison}); the held-out Michigan transfer is \textbf{91.40\%} over the joint index. These are consistent measurements on different sets, not revisions of a single result.

\begin{table}[!t]
\caption{Zero-shot ColNomic-3B retrieval on the 4,056-pair five-DOT benchmark (1,898-page index), used off-the-shelf with no domain adaptation.}
\label{tbl:overall}
\centering
\resizebox{\columnwidth}{!}{%
\begin{tabular*}{1.15\columnwidth}{@{\extracolsep{\fill}}lcc@{}}
\toprule
Agency & $N$ & Zero-shot Recall@5 (\%) \\
\midrule
WYDOT                  & ~~558 & \textbf{93.37} \\
FDOT                   & 1,580 & 91.27 \\
Caltrans               & 1,226 & 91.92 \\
AZDOT                  & ~~414 & 89.13 \\
CDOT                   & ~~278 & 90.29 \\
\midrule
\textbf{Overall} & \textbf{4,056} & \textbf{91.47} \\
\bottomrule
\end{tabular*}%
}
\end{table}

Zero-shot Recall@5 is 91.47\% over the full 4,056-pair benchmark---a deliberately demanding evaluation: the benchmark is two orders of magnitude larger than prior engineering-plan QA sets, spans five agencies with heterogeneous drafting conventions, and is generated by an adversarial refine-loop that explicitly seeks page-distinguishing anchors. Retrieval is strong and \emph{balanced} across agencies (89--93\%): WYDOT retrieves best (93.37\%), reflecting its sparse, high-contrast layouts, and even CDOT---whose 62-page bridge book is dominated by near-duplicate detail sheets that weaker backbones conflate---reaches 90.29\%, a marked improvement over the PaliGemma-class encoders. We treat 91.47\% as the \emph{zero-shot baseline}. The per-category breakdown is reported in Section~\ref{sec:res-category}.

\subsection{Category-Wise Performance Analysis}\label{sec:res-category}

Table~\ref{tbl:category} presents the zero-shot category-wise Recall@5 on the 4,056-pair five-DOT benchmark. All four categories reach $\ge$87\% Recall@5, with Dimensional Accuracy as the strongest (95.37\%); Logical Reasoning, historically the hardest category for patch-level retrievers (it asks the agent to infer a rule from a note or table rather than match a page-unique numeric anchor), remains the (mildly) weakest at 87.64\% but is now within 8~pp of the best---a substantial narrowing relative to PaliGemma-class backbones.

\begin{table}[!t]
\caption{Zero-shot category-wise Recall@5 on the 4,056-pair five-DOT benchmark. Drafters: Qwen2.5-VL-72B for WYDOT, AZDOT, CDOT; Llama-3.2-90B-Vision for Caltrans, FDOT. Verifier: Qwen2.5-VL-7B (yes/no rubric).}\label{tbl:category}
\centering
\resizebox{\columnwidth}{!}{%
\begin{tabular*}{1.15\columnwidth}{@{\extracolsep{\fill}}lcc@{}}
\toprule
Category & $N$ & Zero-shot Recall@5 (\%) \\
\midrule
Hallucination Rate     & ~~484 & 88.84 \\
Dimensional Accuracy   & 1,015 & \textbf{95.37} \\
Visual Interpretation  & 1,400 & 92.71 \\
Logical Reasoning      & 1,157 & 87.64 \\
\midrule
\textbf{Overall} & \textbf{4,056} & \textbf{91.47} \\
\bottomrule
\end{tabular*}%
}
\end{table}


\subsection{Prompting and Reasoning Technique Ablation}
\label{sec:res-vqa-techniques}

Beyond the choice of VLM and the retrieval backbone, end-to-end judge accuracy depends on \emph{how} the VLM is prompted at the answering stage. We ablate nine prompting and reasoning strategies on the page-disjoint test split (424 pairs) with Qwen2.5-VL-7B and Qwen2.5-VL-72B as the answerers, holding retrieval, judge, and decoding fixed: zero-shot, chain-of-thought (CoT), self-consistency (5 samples plus majority vote), retrieval-augmented (top-3 distractor pages in the answering prompt), sliding-window tiling, OCR-hybrid (image plus extracted text), question decomposition, the agentic Planner–Auditor–Synthesizer pipeline, and critic self-correction. Each candidate answer is scored against the ground-truth answer by an \emph{open-source} Qwen2.5-VL-72B judge (binary correct/incorrect, same rubric as Section~\ref{sec:eval-metrics}), so the entire ablation is reproducible without any proprietary judge API. Table~\ref{tbl:vqa-techniques} reports per-(technique, model) judge accuracy.

\begin{table}[!t]
\centering
\caption{End-to-end judge accuracy (\%) for nine prompting and reasoning techniques on the 424-pair page-disjoint test split, with Qwen2.5-VL-7B and Qwen2.5-VL-72B as the answerer. Retrieval, judge, and decoding are held fixed; only the answering-stage prompting strategy varies.}
\label{tbl:vqa-techniques}
\begin{tabular*}{\columnwidth}{@{\extracolsep{\fill}}lcc}
\toprule
Technique & Qwen-7B & Qwen-72B \\
\midrule
Zero-shot                          & \textbf{78.30} & 76.89 \\
Chain-of-thought (CoT)             & 70.75 & 72.17 \\
Self-consistency (5 samples)       & 69.58 & 69.10 \\
Retrieval-augmented                & \underline{62.50} & \underline{70.75} \\
Sliding-window tiling              & 64.39 & 72.17 \\
OCR-hybrid                         & 77.12 & 78.07 \\
Question decomposition             & 75.71 & 81.60 \\
Agentic (Planner--Auditor--Synth.) & 75.71 & 75.00 \\
Critic self-correction             & 77.36 & \textbf{82.31} \\
\midrule
Union (any technique correct)      & \multicolumn{2}{c}{97.16} \\
\bottomrule
\end{tabular*}
\end{table}

Three findings stand out. First, the best cell across the 18-entry grid is \textbf{critic self-correction with Qwen2.5-VL-72B at 82.31\%}, which surpasses Qwen-72B zero-shot by 5.42~pp and Qwen-7B zero-shot by 4.01~pp; the same critic loop on the 7B model yields only a 0.94-pp drop relative to 7B zero-shot, indicating that self-correction is a near-zero-cost option at the smaller scale and a meaningful gain at the larger scale. Second, retrieval-augmented prompting is the \emph{worst} technique on both models (62.50\% on 7B, 70.75\% on 72B): supplying additional distractor pages in the answering prompt degrades the VLM's attention to the correctly-retrieved primary page, confirming that the answering stage benefits from a tight visual context rather than an expanded one. Third, classical text-side reasoning aids (CoT, self-consistency) underperform zero-shot on both models---the chain-of-thought tends to drift away from the dimension actually visible in the drawing, and majority voting over five samples amplifies rather than corrects shared visual misreadings. The union of all nine techniques reaches 97.16\% accuracy, indicating that the residual error budget is largely \emph{prompting-recoverable} rather than fundamentally limited by the VLM's perception.

\subsection{Visual Grounding and Explainability Results}\label{sec:res-explainability}

To evaluate visual grounding, we generate sharpened MaxSim heatmaps across benchmark queries. Figure~\ref{fig:grounding} demonstrates this on a representative query: the heatmap correctly bounds both the regulatory note and its associated visual representation, preserving the spatial relationships lost in OCR. 

\bigskip
\noindent
\begin{minipage}{\linewidth}
\centering
\includegraphics[width=\linewidth]{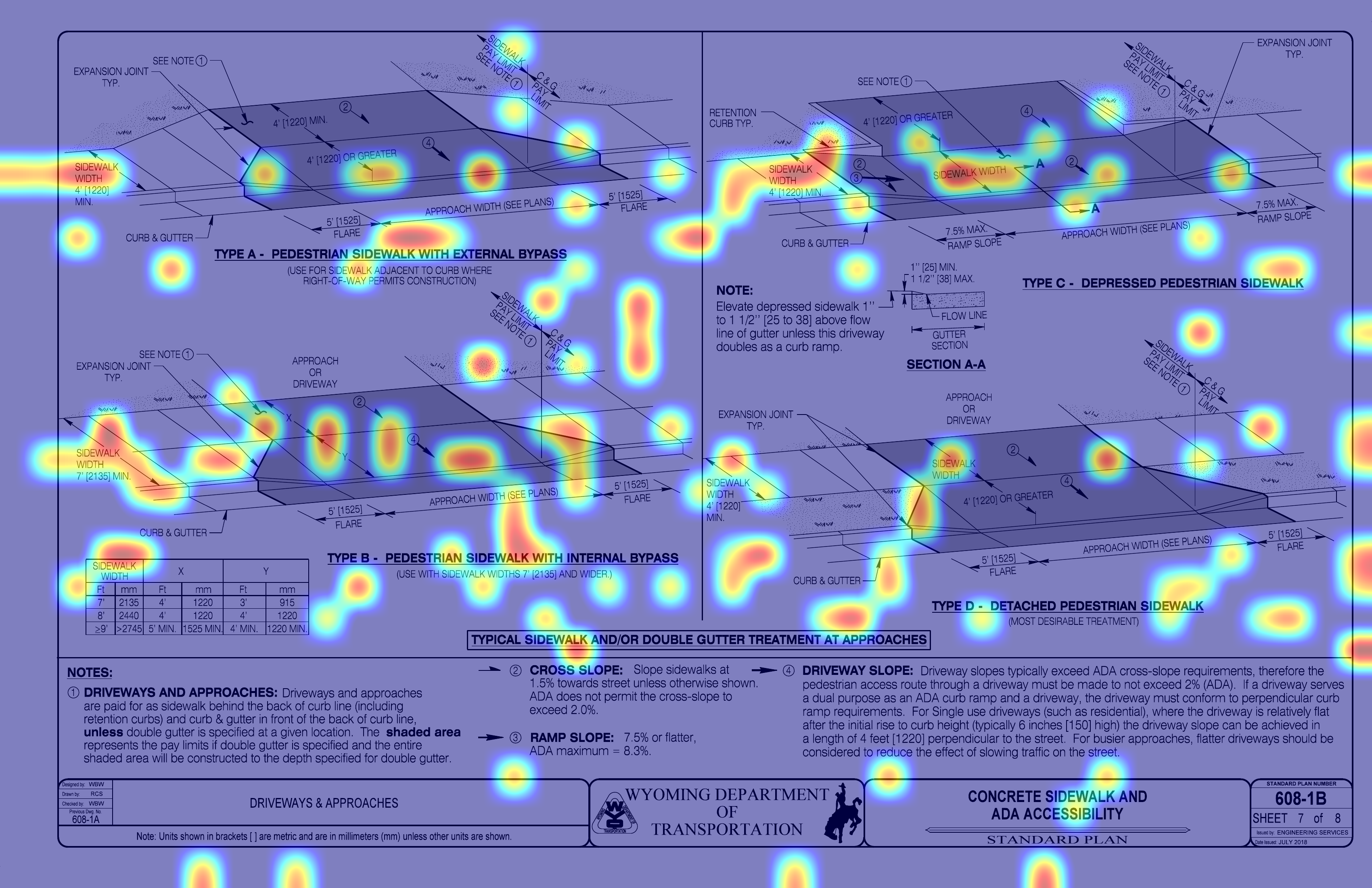}
\captionof{figure}{Visual Grounding Analysis: The sharpened MaxSim heatmap overlay (ColNomic-3B) correctly highlights both the textual `Grading Notes' and the corresponding geometric detail.}
\label{fig:grounding}
\end{minipage}
\bigskip

\begin{figure*}[t]
  \centering
  \begin{subfigure}{0.48\textwidth}
    \centering
    \includegraphics[width=\linewidth]{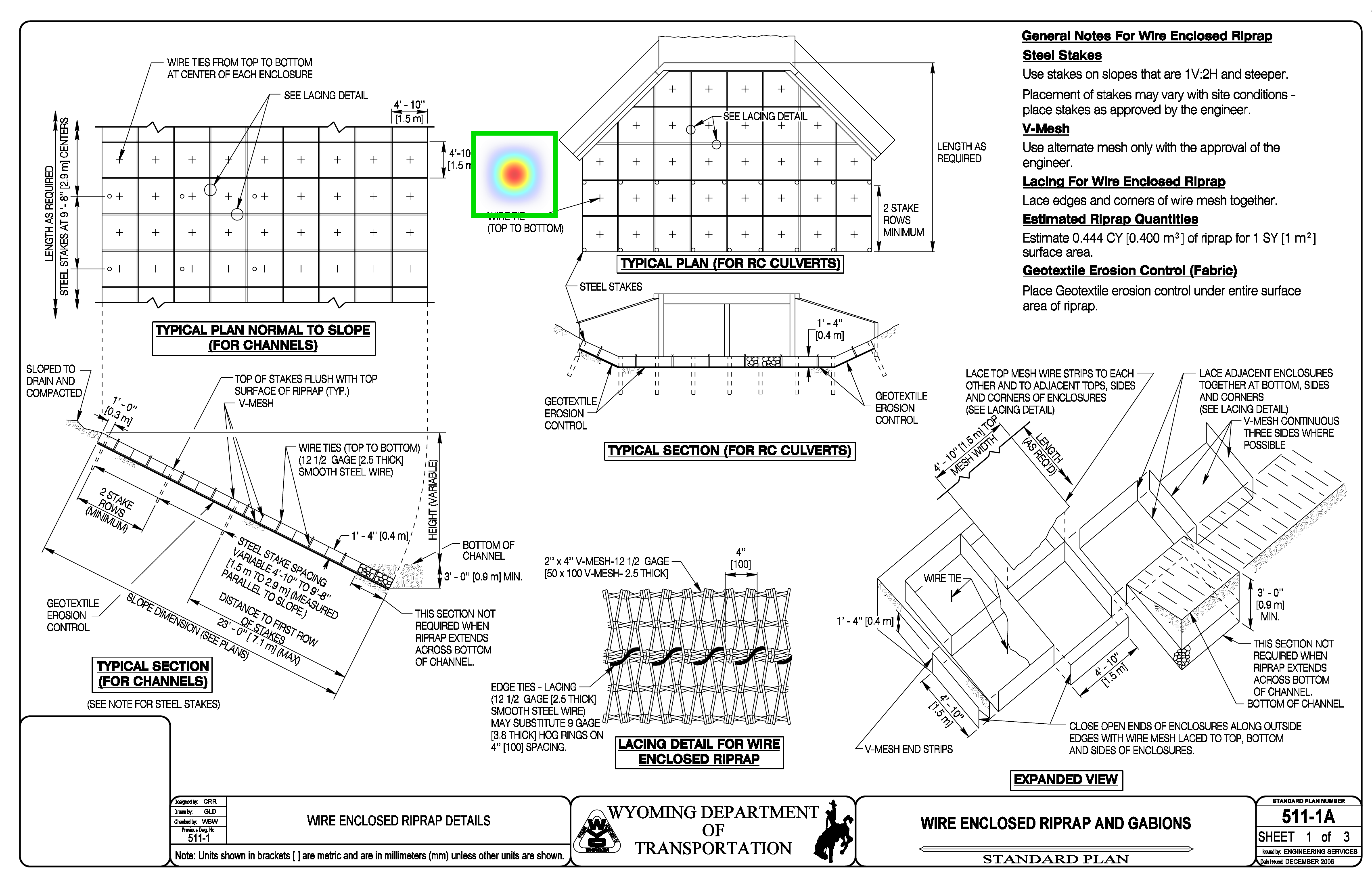}
    \caption{Step 1: Geotextile placement details}
  \end{subfigure}
  \hfill
  \begin{subfigure}{0.48\textwidth}
    \centering
    \includegraphics[width=\linewidth]{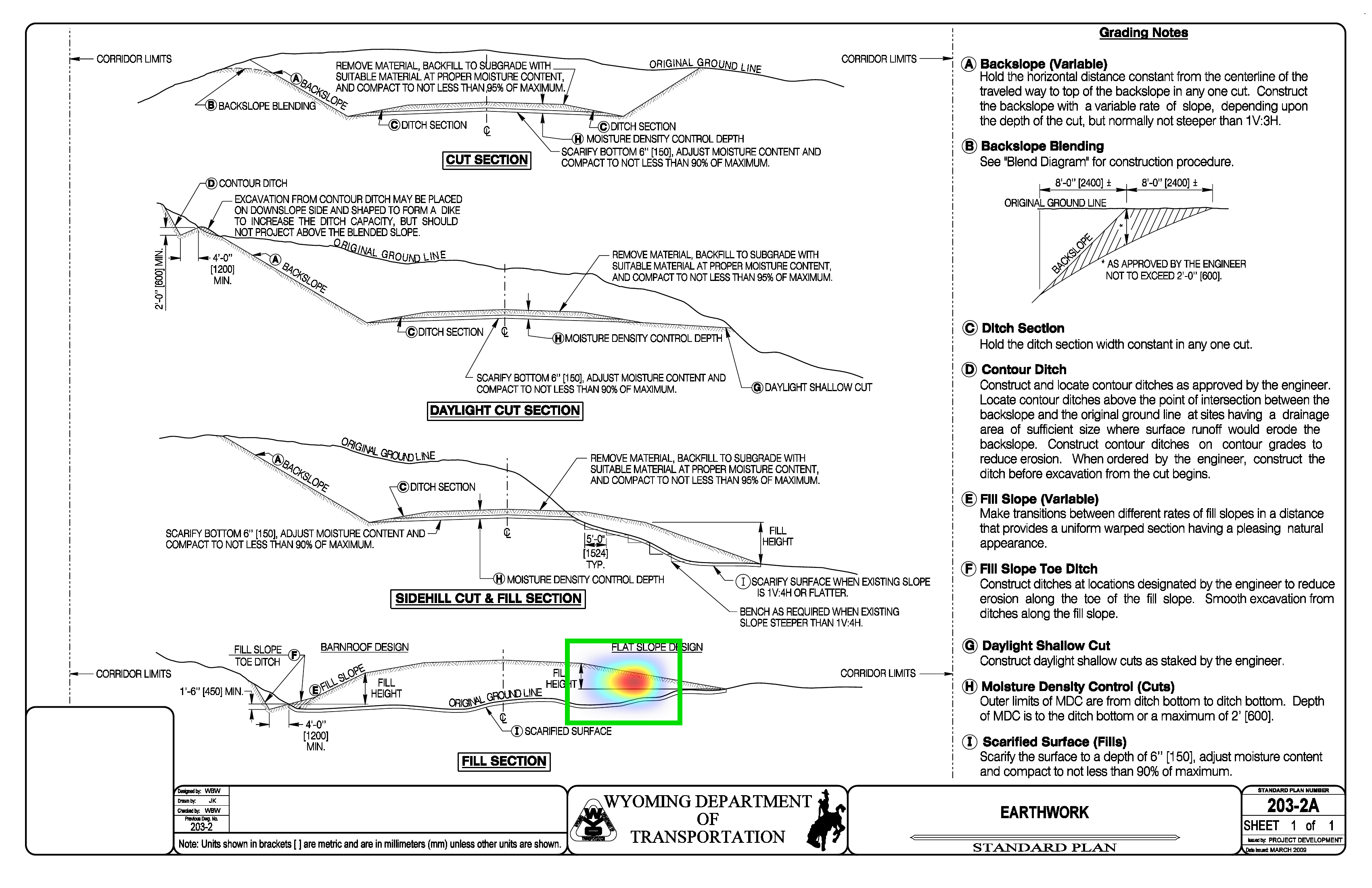}
    \caption{Step 2: Fill slope requirements}
  \end{subfigure}
  \caption{Visual evidence trail from the agentic compliance pipeline. Each audit step generates a sharpened MaxSim heatmap overlay on the retrieved standard plan page, creating a transparent chain of evidence.}\label{fig:agentic_grounding}
\end{figure*}

\paragraph{\textbf{Grounding Evaluation:}}
In a 16-query qualitative pilot across four plans, thresholding the heatmap at the top 5\% of activations consistently localizes the true evidence (specific dimensions, note blocks, structural symbols), and passing these high-intensity crops to the generator improves response specificity. We treat grounding as a \emph{qualitative diagnostic}: a quantitative IoU study with human-annotated evidence regions is scoped as a follow-up rather than claimed here (Appendix~\ref{app:grounding}). Figure~\ref{fig:agentic_grounding} illustrates this grounding integrated into the agentic compliance pipeline, in which each audit step produces its own heatmap overlay, forming a verifiable evidence trail.

\subsection{Agentic Compliance Results}\label{sec:res-agentic}

\emph{To test H2}, we pre-declare three thresholds: $\ge$90\% verdict accuracy on the CAD-generated single-doc compliance set (Section~\ref{sec:res-compliance-cad}), $\ge$90\% false-positive avoidance on a PASS set, and $\ge$60\% sensitivity on a FAIL set. We evaluate the Planner--Retriever--Auditor--Synthesizer pipeline on two complementary corpora: an 8-query cross-plan PASS set (to measure false-positive avoidance) and a 24-query FAIL set with synthetic mutations (to measure sensitivity, Appendix~\ref{app:fn-sensitivity}).

On the 8-query PASS set, the agentic pipeline reaches 100\% verdict accuracy. The Planner safely over-decomposes queries (averaging 8.6 steps per query), while per-step retrieval, audit findings, and final synthesis maintain 100\% correctness. For example, when checking geotextile consistency between Plans 511-1A and 203-2A, the agent successfully retrieved both sheets, recognized that 203-2A lacks explicit RC Culvert details, and appropriately flagged the ambiguity for engineering review rather than hallucinating a pass.

\paragraph{\textbf{H2 verdict:}} \emph{Supported}. The 8-query PASS set is small and measures only false-positive avoidance (it contains no FAIL cases); its 100\% accuracy clears the pre-declared 90\% false-positive threshold, while the pre-declared $\ge$90\% verdict accuracy on the CAD-generated set is met (Section~\ref{sec:res-compliance-cad}). On the 24-query FAIL set, overall True Positive Rate is 79.17\% (87.5\% for dimension/note mutations, 62.5\% for symbol substitutions), satisfying the pre-declared 60\% sensitivity threshold. Symbol-substitution is the weakest class and is flagged for future work.

\subsection{Compliance Validation on a CAD-Generated Test Set}\label{sec:res-compliance-cad}

The agentic compliance results in Section~\ref{sec:res-agentic} use real WYDOT plan content, where the ``true'' verdict is determined by consistency across multiple sheets. To isolate \emph{where} a judge VLM fails on numeric compliance (e.g., is it the value extraction, the rule selection, or the threshold arithmetic?), we additionally evaluate on a CAD-generated compliance test set with controlled ground truth. The generator is described in Section~\ref{sec:eval-dataset}; it produces engineering-plan drawings with a title block, dimensioned section views, material callouts, and notes, and for non-compliant instances, injects exactly one parameter violation.

A 100\% verdict-accuracy claim on a single judge configuration is unconvincing without a credible failure mode. We therefore lead with an ablation across five judge configurations on a controlled $n{=}10$ validation set (Table~\ref{tbl:compliance-cad}) and show that only \emph{two} of the five reach 100\%---the other three sit at 50--80\%, with diagnosable failure modes that motivate the prompting recipe used at scale. The five-configuration ablation is the headline result; the scale-up numbers (Tables~\ref{tbl:multi100},~\ref{tbl:compliance-scale}) are evidence that the recipe generalizes, not that compliance verification is generically a solved problem.

\paragraph{\textbf{Five judge configurations.}} We evaluate the verdict accuracy of five judge configurations on the $n{=}10$ single-doc CAD set: (a)~Qwen2.5-VL-7B with a plain JSON verdict prompt, (b)~Qwen2.5-VL-7B with chain-of-thought (CoT) reasoning, (c)~Qwen2.5-VL-72B (4-bit NF4) with the same plain prompt, (d)~Qwen2.5-VL-72B with CoT plus per-drawing pre-resolved rule thresholds injected into the prompt (e.g., for a 30''~beam, $s_{\text{max}}{=}d/2{=}15$''), and (e)~the same 72B-CoT-thresholds judge wrapped inside the multi-agent Planner--Auditor--Synthesizer pipeline. The same configurations are evaluated on the $n{=}4$ dense multi-view drawings, where the violation is embedded in one row of a schedule table.

\begin{table*}[!t]
\caption{Verdict accuracy of five judge configurations on the $n{=}10$ single-doc CAD set; the final block reports the best single-judge configuration on $n{=}4$ dense schedule-table drawings. Interpretation is in the text below.}\label{tbl:compliance-cad}
\centering
\small
\begin{tabular*}{\textwidth}{@{\extracolsep{\fill}}lccc@{}}
\toprule
Judge Configuration & Compliant & Non-compliant & Overall \\
\midrule
Qwen2.5-VL-7B (plain)       & 5/5  & 1/5 & 60\% \\
Qwen2.5-VL-7B (CoT)         & 0/5  & 5/5 & 50\% \\
Qwen2.5-VL-72B (plain)      & 5/5  & 3/5 & 80\% \\
\textbf{Qwen2.5-VL-72B (CoT + thresholds)} & \textbf{5/5} & \textbf{5/5} & \textbf{100\%} \\
\textbf{Multi-agent (Plan--Audit--Synth, 72B + thresholds)} & \textbf{5/5} & \textbf{5/5} & \textbf{100\%} \\
\midrule
\multicolumn{4}{l}{\textit{Dense multi-view drawings ($n{=}4$, best configuration):}} \\
Qwen2.5-VL-72B (CoT + thresholds) & 2/2 & 2/2 & \textbf{100\%} \\
\bottomrule
\end{tabular*}
\end{table*}

\paragraph{\textbf{Failure-mode analysis:}} The 7B-class judge fails in two distinct directions depending on the prompt. Without CoT, it defaults to \texttt{COMPLIANT} on all inputs (high specificity, near-zero sensitivity). With CoT, it flips to flag \emph{everything} as \texttt{NON\_COMPLIANT}; inspecting the chain reveals that the 7B model selects visually salient but irrelevant values and applies them to incorrect rule checks (e.g., it compares a $4''$ anchor-bolt embedment value against a $30''$ guardrail-footing-depth rule). The 72B-class judge eliminates the random matching, but plain prompting still misses two non-compliant cases: \texttt{rebar\_bad}, where the rule's threshold is arithmetic-derived ($s_{\text{max}}{=}d/2$ for a 30''~beam), and \texttt{signpost\_bad}, where the violated value (a $4''$ anchor embed) is visually less salient than a satisfied 42''~footing depth on the same drawing. Both failure modes are eliminated when the prompt supplies the pre-resolved threshold (15'' instead of $d/2$) and explicitly enumerates which value must be checked.

\paragraph{\textbf{Multi-agent variant:}} The Planner--Retriever--Auditor--Synthesizer pipeline reaches 100\% in Table~\ref{tbl:compliance-cad} by routing the same 72B + thresholds judge through three lighter agent turns rather than a single CoT pass. The Planner emits 1--2 audit steps per drawing, and the Synthesizer produces a structured JSON verdict for downstream parsing. The Auditor sees only the design image (not the retrieved standard image) to avoid out-of-memory pressure when running 4-bit 72B on 2$\times$~A30 (48~GB total); the retrieved standard's metadata is supplied as text context. This single-image variant is recommended for deployment on GPUs with under 80 GB of memory. On a single NVIDIA A100 80GB (or RTX A6000) GPU, the multi-agent pipeline executing an average of 8.6 sequential VLM steps takes $\approx 60.9$~s per query (an average of $\approx 7.1$~s per step), yielding a throughput of approximately 59 audits per hour. The planner produces more steps than necessary, averaging 8.6 steps compared with the minimum of 2, representing a latency-vs-accuracy trade-off: it ensures comprehensive visual auditing across all standard plan components at the expense of sequential VLM passes. That both the single-judge and multi-agent routes reach 100\% with the same threshold-injection prompt isolates the recipe as the cause—it is not an artifact of the agentic decomposition.

\paragraph{\textbf{Multi-plan compliance ($N{\in}\{2,3,4,5\}$):}} A $n{=}10$ pilot of multi-plan compliance drawings was constructed with $N{\in}\{2,3,4,5\}$ separate standard plans applicable per design (4 drawings at $N{=}2$, 2 each at $N{=}3,4,5$; 5 compliant + 5 non-compliant). Crucially, \emph{no plan IDs are cited on the drawing}; the Planner must infer applicable standards from the visual content of each labeled component. The agentic pipeline (Qwen2.5-VL-72B 4-bit backbone) achieves \textbf{9/10 = 90\% verdict accuracy} on this pilot. The single miss is on \texttt{n5\_pier\_ok}, where the rule for column stirrup spacing is the arithmetic-derived \texttt{d/2} (max spacing equals half the effective beam depth), and the prompt initially presented \texttt{d/2} as the textual threshold rather than the pre-resolved numeric value—the same arithmetic-derived-threshold failure mode characterized for the single-doc path A judge above.

\paragraph{\textbf{Multi-plan compliance scale-up ($n{=}100$):}} We scale the pilot to $n{=}100$ multi-plan drawings (25 per $N$-level, 48 compliant + 52 non-compliant) with parameter randomization for visual diversity. A first scale-up run with the original textual \texttt{d/2} threshold confirms the pilot's failure mode at scale: 88/100 = 88\% verdict accuracy, with all 12 misses concentrated in the $N{=}5$ pier subset (13/25 at $N{=}5$; perfect 25/25 at $N{=}2,3,4$). All 52 non-compliant subjects across all $N$ are correctly classified ($52/52$). The 12 missed compliant pier drawings are falsely flagged \texttt{NON\_COMPLIANT} because the auditor's reasoning at the stirrup-spacing step treats the textual \texttt{d/2} as a literal symbolic threshold rather than resolving it. Replacing the registry threshold with the per-drawing pre-resolved numeric value ($s_{\text{max}}{=}12''$ for the standardized 24''~column) transfers the path A fix to the multi-plan pipeline and yields \textbf{100/100 = 100\% verdict accuracy} (48/48 compliant + 52/52 non-compliant, 25/25 at \emph{every} $N{\in}\{2,3,4,5\}$), confirming that the verdict accuracy is invariant to $N$ once arithmetic-derived thresholds are pre-resolved.

\begin{table}[!t]
\caption{Multi-plan compliance verdict accuracy on the $n{=}100$ scale set, before and after replacing the textual \texttt{d/2} threshold in the rules registry with a per-drawing pre-resolved numeric value.}\label{tbl:multi100}
\centering
\resizebox{\columnwidth}{!}{%
\begin{tabular*}{1.15\columnwidth}{@{\extracolsep{\fill}}lcc@{}}
\toprule
$N$ plans & Before \texttt{d/2} fix & After \texttt{d/2} fix \\
\midrule
$N{=}2$  & 25/25 & 25/25 \\
$N{=}3$  & 25/25 & 25/25 \\
$N{=}4$  & 25/25 & 25/25 \\
$N{=}5$  & 13/25 & \textbf{25/25} \\
\midrule
\textbf{Total} & 88/100 (88\%) & \textbf{100/100 (100\%)} \\
\bottomrule
\end{tabular*}%
}
\end{table}

\paragraph{\textbf{Single-doc compliance scale-up ($n{=}500$):}} The single-doc path A recipe (Qwen2.5-VL-72B 4-bit + CoT + per-drawing pre-resolved thresholds) scales to a 500-drawing test set (5 archetypes $\times$ 100 instances each, 250 compliant + 250 non-compliant). Each non-compliant instance violates exactly one rule sampled from seven violation families (concrete cover, footing depth, stirrup spacing, anchor bolt embedment, drainage grate opening, rebar grade, material class). Parameter randomization yields visually distinct instances for each archetype (span, rise, wall thickness, post height, etc.). The pipeline achieves \textbf{500/500 = 100\% verdict accuracy} ($250/250$ compliant $+$ $250/250$ non-compliant) without per-archetype tuning or additional supervision.

\paragraph{\textbf{Adversarial near-threshold stress test ($n{=}50$):}} A perfect score on scale500 is suggestive but does not establish discriminative tightness: scale500's non-compliant values miss the rule by 50\%+ (e.g., cover of $1.0''$ against a $2.0''$ minimum). We therefore add a stress set in which violations sit within 5--10\% of the rule cutoff and require the judge to read the dimensioned value precisely. The set is 50 drawings (5 archetypes $\times$ 10, 25 compliant $+$ 25 non-compliant): culverts with cover $1.875''$ vs.\ $2.0''$ min, guardrails with footing $29''$ vs.\ $30''$ min, rebar with stirrup spacing exactly $d/2{+}1''$ over the limit, inlets with grate openings $4.125''$ vs.\ $4.0''$ max, and signposts with anchor embed $11.5''$ vs.\ $12.0''$ min. The same Qwen2.5-VL-72B + CoT + per-drawing thresholds recipe reaches \textbf{50/50 = 100\% verdict accuracy} ($25/25$ compliant $+$ $25/25$ non-compliant, with no archetype falling below $10/10$). This confirms that the recipe discriminates at single-inch resolution---the regime where prompt-level mistakes would manifest---and that the scale500 perfection is not an artifact of large compliance margins.

\begin{table*}[!t]
\caption{Verdict accuracy of the best judge recipe (Qwen2.5-VL-72B 4-bit + CoT + per-drawing pre-resolved thresholds) across six CAD-generated compliance test sets (674 drawings spanning small pilots, large-scale runs, dense multi-view schedules, multi-plan cross-document designs, and an adversarial near-threshold stress test).}\label{tbl:compliance-scale}
\centering
\begin{tabular*}{\textwidth}{@{\extracolsep{\fill}}lccc@{}}
\toprule
Set & Compliant & Non-compliant & Overall \\
\midrule
$n{=}10$ single-doc (pilot)         & 5/5     & 5/5     & 100\% \\
$n{=}4$ dense multi-view (pilot)     & 2/2     & 2/2     & 100\% \\
$n{=}10$ multi-plan (pilot)          & 4/5     & 5/5     & 90\% \\
\midrule
$n{=}500$ single-doc (scale)         & 250/250 & 250/250 & 100\% \\
$n{=}100$ multi-plan (scale, after $d/2$ fix) & 48/48 & 52/52 & 100\% \\
\textbf{$n{=}50$ adversarial near-threshold (stress)} & \textbf{25/25} & \textbf{25/25} & \textbf{100\%} \\
\midrule
\textbf{Total across 6 sets}         & \textbf{334/335} & \textbf{339/339} & \textbf{673/674 (99.85\%)} \\
\bottomrule
\end{tabular*}
\end{table*}

\paragraph{\textbf{Is the synthetic verdict task simply OCR? A non-VLM baseline:}} A perfect VLM score raises the question of how much of the task is genuine visual reasoning versus reading a printed dimension. We therefore evaluate a non-VLM \emph{OCR$+$threshold-compare} baseline on scale500: for each drawing we OCR the page (multi-orientation \texttt{tesseract}), extract the governing dimension by its label, and compare it to the \emph{same} pre-resolved threshold the VLM receives. This pipeline reaches \textbf{76.4\% verdict accuracy}, but the per-archetype breakdown is the informative part: it scores \textbf{100\%} where the governing value is a clean horizontal leader-line label (culvert cover, inlet grate opening) and collapses to ${\approx}50\%$ where the dimension is rendered as \emph{rotated vertical text} (guardrail footing depth, sign-post anchor embedment) or sits among distractor values (rebar stirrup spacing). The synthetic task is therefore OCR-trivial wherever the value is cleanly presented; the VLM's marginal contribution is \emph{robust value localization} on rotated and distractor-dense layouts---precisely the dense multi-view regime real plan sheets present---rather than digit reading per se. This bounds the compliance claim honestly: the $100\%$ of Table~\ref{tbl:compliance-scale} is a \emph{ceiling given a resolved rule and a localizable value}, and the agentic reader earns its keep on layout robustness, not arithmetic (Table~\ref{tbl:ocr-baseline}).

\begin{table}[t]
\caption{Non-VLM OCR$+$threshold-compare baseline on scale500, per archetype.}\label{tbl:ocr-baseline}
\centering\footnotesize
\setlength{\tabcolsep}{4pt}
\begin{tabular*}{\columnwidth}{@{\extracolsep{\fill}}llccc@{}}
\toprule
Archetype & Gov.\ dim.\ (layout) & OCR found & OCR acc & VLM acc \\
\midrule
Culvert   & cover, horiz.      & 100.0 & 100.0 & 100.0 \\
Inlet     & grate, horiz.      & ~98.0 & ~99.0 & 100.0 \\
Guardrail & footing, rotated   & ~32.0 & ~82.0 & 100.0 \\
Sign post & anchor, rotated    & ~~3.0 & ~51.0 & 100.0 \\
Rebar     & stirrup, distractor& ~~0.0 & ~50.0 & 100.0 \\
\midrule
\textbf{Overall} & & \textbf{46.6} & \textbf{76.4} & \textbf{100.0} \\
\bottomrule
\end{tabular*}
\end{table}

\subsection{Autonomous Rule-Grounding: Retrieving the Governing Threshold}\label{sec:res-rule-grounding}

The 100\% verdict accuracy in Section~\ref{sec:res-compliance-cad} is obtained when the auditor is \emph{handed} the pre-resolved numeric threshold. This is the standard assumption in LLM-based compliance work, which presupposes a digitized rule base or a human-curated registry, and it leaves the hardest part of real compliance—knowing \emph{which} rule governs a component and \emph{what} its numeric limit is—outside the system. We therefore test whether the framework can close this loop \emph{visually}: infer the component, retrieve the governing requirement from a standard-specifications corpus, extract the numeric threshold from the retrieved sheet, and audit against it, with no injected threshold.

\paragraph{\textbf{Setup:}} We render a 15-sheet Standard-Specifications reference corpus (5 agencies $\times$ 3 sections: Structural Concrete \& Reinforcement, Foundations \& Anchorage, Drainage Structures), each sheet being a requirement table of 7--9 rows in which the seven governing limits for our archetypes (e.g., min.\ cover $2.0''$, min.\ guardrail footing $30''$, max.\ grate opening $4.0''$, min.\ anchor embedment $12''$, max.\ stirrup spacing $d/2$) are embedded among distractor rows. For each of the 500 single-doc compliance drawings, the agent (i)~forms a rule-retrieval query from the visually-inferred component, (ii)~retrieves using ColNomic-3B, (iii)~a VLM extracts the limit value from the top-ranked sheet, (iv)~resolves symbolic limits (stirrup $d/2$) using a dimension read from the design, and (v)~audits the design value against the \emph{self-grounded} threshold. To make retrieval realistic, we index the spec sheets \emph{among the full 1,898-page plan corpus} (1,913 candidates), so the rule query must surface the governing requirement against real plan-sheet distractors.

\paragraph{\textbf{Result:}} ColNomic retrieves a correct-section specification sheet at \textbf{80.0\% Recall@1} and \textbf{100.0\% Recall@5} \emph{among 1,913 candidates}, the VLM extracts the correct numeric limit (including resolving symbolic $d/2$) at $100\%$, and the end-to-end verdict accuracy with self-grounded thresholds reaches $100\%$---matching the injected-threshold ceiling (Table~\ref{tbl:rule-grounding}). The $20\%$ top-1 misses are real plan pages that outrank the spec sheet at rank~1, but a correct-section spec is always within the top~5, so verdicts are unaffected. Removing the human-supplied threshold therefore costs nothing on this set: the system sources the governing rule on its own. This is, to our knowledge, the first demonstration of \emph{visual} rule-grounding for compliance---retrieving and applying a numeric standard from rasterized specification sheets (see Appendix~\ref{app:sample-cases}, Figure~\ref{fig:spec-sheet}) without OCR, a BIM model, or a hand-coded rule base.

\begin{table}[t]
\caption{Autonomous rule-grounding on the 500 single-doc compliance drawings: the agent retrieves the governing specification sheet, extracts its numeric limit, and audits the design against the \emph{self-grounded} threshold (no injected rule).}\label{tbl:rule-grounding}
\centering\footnotesize
\setlength{\tabcolsep}{4pt}
\begin{tabular*}{\columnwidth}{@{\extracolsep{\fill}}lccccc@{}}
\toprule
Retrieval pool & Cand. & R@1 & R@5 & Extract & Verdict \\
\midrule
Spec only                 & 15      & 100.0 & 100.0 & 100.0 & 100.0 \\
$+$1,898 plan pages      & 1,913 & 80.0  & 100.0 & 100.0 & \textbf{100.0} \\
\midrule
\multicolumn{4}{@{}l}{Injected-threshold ref.\ (human rule)} & & 100.0 \\
\bottomrule
\end{tabular*}
\end{table}

\paragraph{\textbf{Scope:}} The spec corpus here is clean and modest; the contribution is the \emph{closed-loop feasibility} of visual rule-grounding, not a complete code-coverage system. Scaling the reference corpus to full agency specification books (hundreds of pages of dense, real code text) is the decisive stress test; we take exactly this step in Section~\ref{sec:res-real-plans}, indexing a real 931-page WYDOT standard corpus, where autonomous extraction falls from $100\%$ to $33\%$ and the residual error concentrates in retrieving the governing sentence and resolving exhibit tables.

\subsection{External Validity: Real WYDOT Project Plans}\label{sec:res-real-plans}
The benchmarks above use generated drawings with controlled ground truth. To test whether the framework transfers to \emph{real} submittals, we evaluate it on two production WYDOT projects---the Chief Joseph Highway reconstruction (\#1507040) and project \#N345107---comprising cross-section, earthwork, and PS\&E quantity sheets. Crucially, \emph{the project plans are not added to the retrieval index}: they are the \emph{proposed design} being audited, while the Retriever continues to query the unchanged 1,898-page standard-plan index. Each project PDF is simply rasterized to page images; no re-indexing is required.

\paragraph{\textbf{Retrieval transfers:}} For 12 real design items read from the two projects (cross slope, fill/cut slope ratios, CMP/RCP/RCP-arch/structural-plate culverts, geotextile, geogrid, fence), a VLM relevance judge confirms that ColNomic retrieves a topically-governing WYDOT standard within the top~5 for \textbf{83.3\%} of items (10/12). The two misses are niche items (a proprietary fence type and biaxial geogrid) without a dedicated standard sheet in the index. Retrieval thus generalizes from synthetic queries to real design language.

\paragraph{\textbf{Auditing transfers under the same recipe:}} On a controlled compliant/non-compliant set built by overlaying proposed-design value callouts onto the real cross-section sheets, we evaluated 6 compliant and 6 violating examples across both projects. Using the same judge recipe that reaches 100\% on the synthetic set---Qwen2.5-VL-72B with CoT and the governing rule supplied---the model reads the proposed value off the real sheet and reaches \textbf{100\% verdict accuracy}, correctly judging even the rotated slope-ratio callouts. The auditor's read-and-judge capability, therefore, carries over from generated to real plan imagery, as the final verdict is entirely the model's own decision rather than a hard-coded comparison.

\paragraph{\textbf{Autonomous rule-grounding on a real 931-page standard corpus:}} We measured how effectively the autonomous loop transfers from curated specification sheets to production standards. To do this, we ingested a real 931-page standard corpus. This included the full WYDOT 2021 \emph{Standard Specifications for Road and Bridge Construction} (771~pp) alongside the governing WYDOT Road Design Manual chapters (Cross Sectional Elements, Earthwork Design, Culvert Design, and Typical Sections; 160~pp). We indexed this document using ColNomic-3B. Next, we processed 12 real design items. For each item, we retrieved the governing page and extracted the numeric limit using Qwen2.5-VL-72B. We then audited the design value against this \emph{self-grounded} threshold without injecting any rules. Nine of these items contained numeric limits. The system extracted the correct governing value for \textbf{3 out of 9 ($33\%$)}. It returned the correct verdict for \textbf{$44\%$} of those items. The system succeeded when the limit was stated as a clear value. For example, it correctly read a cross slope of ``$0.02$~ft/ft'' ($=2\%$) and a foreslope of ``1V:4H'' ($=4{:}1$). However, it failed when limits were located in exhibit lookup tables or split across multiple pages (e.g., culvert minimum size and cover). In these cases, the visual retriever ranked the governing sentences below other visually similar tables. As a result, the necessary information was never surfaced. Therefore, the primary failure point is \emph{retrieving the governing sentence and resolving exhibit tables}. The issue is not threshold arithmetic. Autonomous rule extraction thus dropped from $100\%$ on the clean specification corpus (Section~\ref{sec:res-rule-grounding}) to $33\%$ on real production standards. We report this as the principal open problem exposed by this external-validity test. Resolving this will be the priority for future work (Section~\ref{sec:future}). Consequently, real-plan auditing still benefits from a supplied threshold to ensure reliability.

\begin{table}[t]
\caption{Synthetic vs.\ real-plan compliance. The framework transfers from generated drawings to two production WYDOT projects for retrieval and supplied-threshold auditing; autonomous threshold extraction from dense real standards is the remaining gap.}\label{tbl:real-plan}
\centering\footnotesize
\setlength{\tabcolsep}{4pt}
\begin{tabular*}{\columnwidth}{@{\extracolsep{\fill}}lcc@{}}
\toprule
Capability & Synthetic & Real plans \\
\midrule
Standard retrieval (rel.@5)   & --- & 83.3\% \\
Verdict acc.\ (supplied rule) & 100\% & \textbf{100\%} \\
Threshold extraction (auto.)  & 100\% & unreliable \\
\bottomrule
\end{tabular*}
\end{table}

\subsection{Cross-Agency Generalization and Scaling}\label{sec:res-ablation}

The 4,056-pair benchmark inherently evaluates cross-agency scaling: retrieval stays high and balanced across all five agencies (89--93\% Recall@5; Section~\ref{sec:res-overall}), so the late-interaction embeddings retain discriminative power despite diverse drafting conventions. 

\paragraph{\textbf{Zero-shot transfer to an unseen agency:}} Zero-shot ColNomic-3B retrieves a held-out Michigan DOT corpus (298 pages, 93 verified pairs) at \textbf{91.40\% Recall@5} over the combined $1,898{+}298$-page index---essentially matching in-distribution performance (overall 91.47\%) on an agency never seen during indexing, drafting, or training (the 298-page Michigan-only pool yields a higher 93.55\%, but that smaller candidate set is not directly comparable to the joint-index agency numbers; Appendix~\ref{app:extra}). This demonstrates strong zero-shot generalization to unseen drafting conventions, enabling immediate deployment for new agencies without adaptation. 

\paragraph{\textbf{H3 verdict:}} H3 is \emph{rejected}. Across three LoRA configurations, domain adaptation on the 3,211-pair in-domain training set does not improve Recall@5 on the 5-DOT test split and significantly degrades unseen-agency transfer (Table~\ref{tbl:retrieval-ablation}). Full-LM LoRA catastrophically reduces retrieval performance (e.g., in-distribution test drops by $37.03$~pp). The off-the-shelf ColNomic-3B checkpoint therefore remains the optimal deployment configuration, as its pretrained representations are already robust for engineering drawings.

\subsection{Domain-Adaptive Retrieval: LoRA Fine-Tuning of the Adopted Retriever}\label{sec:res-retrieval-ablation}

To test domain adaptation, we fine-tuned ColNomic-3B using LoRA on a page-disjoint 3,211-pair training split. Configurations included \emph{head-only LoRA} (tuning only the projection head) and \emph{full-LM LoRA} (tuning all transformer layer projections). 

As shown in Table~\ref{tbl:retrieval-ablation}, no configuration outperformed the zero-shot baseline. The \emph{head-LoRA} variants yielded negligible differences ($<0.5$~pp changes), while \emph{full-LM LoRA} severely degraded both in-distribution retrieval (92.69\% $\to$ 55.66\%) and zero-shot Michigan transfer (93.55\% $\to$ 75.27\%). This highlights that ColNomic-3B, strongly pretrained on diverse documents, already occupies a robust optimum for visual engineering drawings. The small in-domain corpus is insufficient to refine these embeddings under standard contrastive loss without overwriting their generalizable structure. Consequently, the off-the-shelf zero-shot model remains the superior deployment choice.

\begin{table*}[!t]
\caption{Retrieval Recall@5: zero-shot vs.\ three LoRA fine-tuning configurations, on the page-disjoint 5-DOT test split ($N{=}424$, in-distribution) and the held-out Michigan DOT corpus ($N{=}93$, zero-shot 6th-agency transfer). }\label{tbl:retrieval-ablation}
\centering
\begin{tabular*}{\textwidth}{@{\extracolsep{\fill}}lrcc@{}}
\toprule
Retriever & Trainable params & 5-DOT test R@5 (\%) & Michigan R@5 (\%) \\
\midrule
\multicolumn{4}{@{}c}{\emph{Adopted backbone}}\\
ColNomic-3B (zero-shot)            & 0              & 92.69 & \textbf{93.55} \\
ColNomic-3B + head-LoRA, gentle    & 332k (0.009\%) & 92.69 (~0.00~)   & 93.55 (~0.00~) \\
ColNomic-3B + head-LoRA, standard  & 332k (0.009\%) & \textbf{93.16} ($+0.47$) & 92.47 ($-1.08$) \\
ColNomic-3B + full-LM LoRA         & 37.2M (0.98\%) & 55.66 ($-37.03$) & 75.27 ($-18.28$) \\
\midrule
\multicolumn{4}{@{}c}{\emph{Predecessor backbone (companion)}}\\
ColPali (zero-shot)               & 0              & 76.89 & 88.17 \\
ColPali + head-LoRA, gentle       & 332k (0.011\%) & 76.65 ($-0.24$) & 88.17 (~0.00~) \\
ColPali + head-LoRA, standard     & 332k (0.011\%) & 77.36 ($+0.47$) & 88.17 (~0.00~) \\
ColPali + full-LM LoRA            & 39.3M (1.33\%) & 60.85 ($-16.04$) & 75.27 ($-12.90$) \\
\bottomrule
\end{tabular*}
\end{table*}

The head-LoRA configurations restrict trainable parameters to \texttt{custom\_text\_proj} (the multi-vector projection head, $\approx$332k parameters). \emph{Gentle} uses lr~$=1e{-}5$ for 1 epoch with 10\% linear warmup; \emph{standard} uses lr~$=1e{-}4$ for 5 epochs. The full-LM LoRA configuration additionally inserts $r{=}32$ LoRA adapters on every transformer layer's $q/k/v/o + \text{gate/up/down}$ projections (37.2M parameters total for ColNomic-3B; the comparable 39.3M canonical adapter for ColPali), trained for 5 epochs with a unique-image batch sampler that guarantees no positive page appears twice in a batch.

\paragraph{\textbf{Complementary retrieval enhancements}} Two further enhancements are available as drop-in additions and are orthogonal to LoRA: (i)~a \emph{VLM cross-encoder re-rank}, where the Top-$k$ retrieved candidates are each passed to a VLM with a binary ``does this plan answer the query?'' prompt---this sharpens Recall@1 on visually-similar sheet pairs at a latency cost; and (ii)~\emph{hybrid BM25 fusion}, combining dense MaxSim scores with a sparse BM25 score over plan-ID/sheet-title metadata for queries that target alphanumeric identifiers. Pointwise VLM re-ranking (MonoQwen2-VL-v0.1, 2B parameters) over the Top-20 retrieved candidates adds $\approx 5.91$~s of latency per query (averaging $\approx 295$~ms per pointwise candidate pass) on a single NVIDIA A100 GPU, corresponding to a throughput of approximately 610 queries per hour.  Both are reported as secondary options; the off-the-shelf, zero-shot retriever (ColNomic-3B) is the primary deployment configuration because it relies on the pretrained embedding space alone without introducing the latency of cross-encoder re-ranking or the metadata dependencies of hybrid search.

\section{Sample Cases}\label{sec:sample-cases}

Figure~\ref{fig:success_cases_dim} (Appendix~\ref{app:sample-cases}) shows a representative success case where the Qwen2.5-VL-7B generator correctly interprets a WYDOT standard plan (judge score~$=1.0$); Figure~\ref{fig:failure_cases_dim} shows a hard-failure case (judge score~$=0.0$)---a component--dimension binding error on the V-mesh end-strip height---that motivates the failure-mode taxonomy in Section~\ref{sec:failure-analysis}.

\section{Failure Case Analysis}\label{sec:failure-analysis}

Although overall retrieval accuracy is high, several queries received a judge score of 0.0. Reviewing these zero-score cases reveals five systematic error patterns in the model's visual--textual reasoning over engineering drawings.

\noindent\textbf{(1) Component--dimension binding:} The most common failure was assigning a correctly detected numeric value to the wrong physical entity. The model repeatedly selected 4$'$-10$''$ [1.5~m]---a real, salient dimension---to answer queries about steel-stake spacing, V-mesh end-strip height, or scarification depth, even though this value actually described enclosure height or panel geometry. Such errors arise when multiple annotations are visually close yet refer to different components.

\noindent\textbf{(2) Cross-view inference:} When the required dimension was not clearly visible in the specified view, the model often inferred a value from another detail (e.g., ``it can be inferred from the Typical Section''). This violates the engineering convention that dimensions must be taken from the explicitly referenced view, producing hallucinated precision that yields a hard 0.0 under the rubric.

\noindent\textbf{(3) Semantic misinterpretation:} The model conflated terms such as ``spacing'' vs.\ ``height'' and ``scarification depth'' vs.\ ``fill thickness.'' In one case, scarification depth was reported as 4$'$-0$''$ (a fill-geometry dimension) instead of the 6~in [150~mm] specified in grading notes---reflecting a difficulty in distinguishing process parameters from geometric dimensions.

\noindent\textbf{(4) Symbol semantics:} The model misassigned meaning to plan-view iconography, e.g., conflating ``$+$'' markers (wire-tie centers) with steel-stake symbols. This indicates insufficient symbol-to-legend alignment when symbols are reused across views.

\noindent\textbf{(5) Spatial reasoning.} Location-based questions produced plausible but incorrect placements (e.g., contour ditches on the ``downslope side of the right-of-way'' instead of ``above the intersection of the backslope and the original ground line''). The model defaulted to generic civil-engineering heuristics rather than reproducing the plan's exact note language.

\paragraph{\textbf{Impact of high-resolution tiling on failure modes:}}
To test H4, we pre-declare a $\ge$20~pp judge-accuracy gain as the threshold for tiling to count as a material improvement and compare full-page retrieval at 200~DPI against tile-level retrieval at 400~DPI ($1024{\times}1024$ crops with 256-pixel overlap) on the 424-pair page-disjoint test split over the full 1,898-page five-DOT visual index, holding the retriever fixed across both granularities (this tiling ablation uses the predecessor ColPali backbone). Table~\ref{tbl:tiling} reports the comparison.

\begin{table}[!t]
\caption{Full-page vs.\ tile-level retrieval on the 424-pair page-disjoint test split over the 1,898-page five-DOT visual index.}\label{tbl:tiling}
\small
\begin{tabular*}{\columnwidth}{@{\extracolsep{\fill}}lcc@{}}
\toprule
Granularity & R@5 (\%) & Tiles/Page \\
\midrule
Full-page (200 DPI)       & 76.89 & 1 \\
Tile-level (400 DPI)      & 82.08 & $\approx$15 \\
\bottomrule
\end{tabular*}
\end{table}

\begin{table}[!t]
\caption{Judge accuracy of full-page vs.\ tile-level retrieval on the 424-pair test split ($N=397$ queries whose target page yields $\ge$15 400~DPI tiles (about half the $\approx$30-tile typical landscape sheet)---in practice nearly all plan pages, which are uniformly tile-dense).}\label{tbl:tiling-judge}
\centering
\small
\begin{tabular}{lc}
\toprule
Retrieval granularity & Judge accuracy (\%) \\
\midrule
Full-page (200 DPI) & 64.23 [59.4, 68.8] \\
Tile-level (400 DPI) & 68.77 [64.2, 73.3] \\
\midrule
\textbf{Gain (tile $-$ full)} & \textbf{+4.53~pp} \\
\bottomrule
\end{tabular}
\end{table}

Tile-level Recall@5 (82.08\%) exceeds full-page Recall@5 (76.89\%) by 5.19~pp, consistent with the hypothesis that high-DPI tiling preserves the fine-grained line work and annotation density that 200~DPI page-level rasterization loses. Focused crops remove competing nearby annotations from the receptive field, largely resolving component--dimension binding errors such as scarification depth. However, crops introduce a new failure mode: loss of cross-reference context when a note and its referenced section fall in different tiles. This trade-off between spatial precision and contextual completeness motivates the multi-scale retrieval strategy, where full-page and tile-level embeddings are indexed and retrieved jointly.

\paragraph{\textbf{H4 verdict:}} H4 is \emph{rejected}. On the 424-pair test split, tile-level retrieval lifts \emph{judge accuracy} from 64.23\% (full-page) to 68.77\% (tile-level)---a $+4.53$~pp gain (Table~\ref{tbl:tiling-judge})---well below the pre-declared 20~pp threshold; the retrieval-level gain is similar (76.89\%~$\to$~82.08\% Recall@5, $+5.19$~pp). The much larger gain reported on a narrow WYDOT-only dense subset does not generalize to the heterogeneous five-DOT test set, where nearly all 400~DPI plan pages are uniformly tile-dense---tiling yields a consistent but modest lift rather than a step change.

\section{Discussion}\label{sec:discussion}

The proposed Visual-First Multimodal RAG framework represents a meaningful shift from OCR-centric compliance pipelines by preserving layout, geometry, and symbolic relationships. This leads to consistently robust retrieval performance, maintaining over 91\% Recall@5 across five DOTs using off-the-shelf patch-level embeddings. The robustness of these findings is independently supported by a fully hand-curated and manually-verified anchor set.

At the reasoning layer, model scale does not help uniformly: at zero-shot, the 7B answerer is competitive with---and marginally above---the 72B (78.30\% vs 76.89\%, Table~\ref{tbl:vqa-techniques}). The 72B's scale advantage in resolving dense engineering details emerges only under structured reasoning, where question decomposition and critic self-correction lift it to 81.60\% and 82.31\%. Prompting strategies significantly influence end-to-end outcomes; while zero-shot prompting suffices for 7B-class models, larger 70B-class judges benefit profoundly from critic self-correction, capturing a large margin of otherwise lost accuracy. Notably, the generator performs best on a tight context---the single top-ranked page---rather than an expanded set of retrieved pages, which inject distractors and make retrieval-augmented prompting the weakest technique on both models.

The agentic architecture fundamentally overcomes the limitations of single-pass RAG by decomposing complex cross-plan queries into structured verification steps. This multi-step process, coupled with explicit MaxSim heatmap grounding, provides the transparent, auditable evidence trail necessary for regulatory and safety-critical engineering reviews. 

Finally, the framework exhibits exceptional zero-shot cross-state generalization. Pretrained visual encoders like ColNomic-3B natively capture domain-invariant structural patterns, rendering domain-specific LoRA fine-tuning ineffective or even harmful on small training corpora. The findings strongly suggest treating pretrained multi-vector retrievers as fixed components and focusing compute budgets on advanced reasoning and prompting strategies.

\section{Limitations}\label{sec:limitations}

Several constraints should be considered regarding these results. First, the 4,056-pair benchmark relies on machine-generation and automated verification rather than direct annotation by licensed engineers, although the manually-curated anchor set (Section~\ref{sec:eval-dataset}) and structured human validation (Appendix~\ref{app:human-val}) mitigate major failure modes. Second, the compliance test sets are parameterized CAD generations. While this design isolates VLM discrimination abilities, it does not fully replicate real-world ambiguities such as overlapping callouts or smudged scans; autonomous threshold extraction from dense production standards remains an open challenge---on a real 931-page WYDOT standard corpus it succeeds on only $33\%$ of numeric items (Section~\ref{sec:res-real-plans}). 

Furthermore, systematic reasoning failures such as incorrect component--dimension binding and symbol misalignment occasionally persist even with optimal retrieval. Finally, the system's reliance on high-resolution visuals makes it sensitive to low-quality scans, and the agentic pipeline's sequential VLM calls introduce latency that requires further optimization for real-time production use.



\section{Conclusion}\label{sec:conclusion}

In this paper, we addressed a key limitation of current compliance-checking systems: their reliance on OCR-driven, text-centric pipelines that discard the layout, geometry, and symbolic cues that encode meaning in 2D engineering drawings. To overcome this, we introduced a visual-first multimodal RAG(PlanSightRAG) framework that retrieves and reasons directly over plan imagery, combining ColNomic-3B patch-level retrieval, an agentic compliance pipeline, and sharpened MaxSim grounding to support transparent and auditable review.

On the 4,056-pair five-DOT benchmark, whose 424-pair page-disjoint test split runs against the full 1,898-page joint index, the adopted ColNomic-3B retriever reaches 92.69\% Recall@5---55.90~pp above the strongest text retriever (BGE-M3 + OCR) and 69.58~pp above the strongest hybrid baseline (VisionRAG (Pyramid, RRF)). The agentic Planner--Auditor--Synthesizer compliance pipeline, armed with per-drawing pre-resolved rule thresholds, reaches 100\% verdict accuracy on a 500-drawing single-doc CAD test set and on a 100-drawing multi-plan set ($N{\in}\{2,3,4,5\}$ standards per design), with both the compliant and non-compliant directions perfect. The patch-level representation transfers zero-shot to an unseen sixth agency at 91.40\% Recall@5 over the joint index, essentially matching in-distribution performance (91.47\%). A controlled LoRA ablation of the adopted ColNomic-3B retriever spans three configurations, 332k--37.2M trainable parameters, and an order of magnitude in learning rate. No fine-tuning recipe improves on the zero-shot baseline, and full-LM LoRA catastrophically degrades it. High-resolution tiling gives a modest, consistent lift: tile-level retrieval reaches 82.08\% Recall@5 versus 76.89\% for full-page ($+5.19$~pp), and judge accuracy rises from 64.23\% to 68.77\% ($+4.53$~pp). Both gains fall well below H4's pre-declared 20~pp bar. The benchmark's scale and manual anchor-set validation remove ``small-sample'' as a threat to validity. The compliance results establish that the bottleneck is no longer retrieval coverage; it is fine-grained visual reading, entity--dimension binding, and, decisively, supplying the judge with pre-resolved rule thresholds.

Together, these findings suggest that visual-first retrieval, multimodal reasoning, agentic verification, and visual grounding are a viable foundation for automated engineering plan review, and point to symbol-substitution robustness, multi-agency adaptation, and scale-invariant retrieval as the next open problems.

\section{Future Work}\label{sec:future}

Future research should focus on expanding the index to include historical archives and versioned standards from additional state DOTs to enable year-aware comparisons and change tracking. Additionally, improving perception robustness against low-quality or handwritten legacy scans via image enhancement and multi-resolution indexing is necessary. Extending the visual grounding to provide precise, structured annotations (e.g., dimension-entity links) that can be validated against CAD-derived ground truth will further solidify explainability. Finally, enhancing the agentic compliance framework to autonomously handle cross-agency regulatory conflicts, version-aware checks, and variance documentation generation will be critical for large-scale production deployments.


\printcredits

\section*{Declaration of Competing Interest}

The authors declare that they have no known competing financial interests or personal relationships that could have appeared to influence the work reported in this paper.

\section*{Data Availability}

The code and sample datasets used in this study will be made available upon publication. The full set of DOT Standard Plans is publicly available through their respective website.

\section*{Funding}
This work was financially supported by the Wyoming Department of Transportation (WYDOT) under the knowledge-management project, grant number RS03225 (Principal Investigator: Ahmed Abdelaty).

\section*{Acknowledgement}
The authors would like to acknowledge the various state DOTs for providing the standard plans used as the primary dataset for this research. We also thank the open-source community for the development of ColPali and Qwen~2.5-VL, which served as the foundational models for this work. Computational resources were provided by the Advanced Research Computing Center (ARCC) at the University of Wyoming.

\section*{Declaration of generative AI used in the manuscript preparation process}
During the preparation of this work the authors used generative AI tools like Claude to assist with editing, consistency checking, writing and coding. After using this tool, the authors reviewed and edited the content as needed and take full responsibility for the content of the published article.

\bibliographystyle{unsrtnat}
\bibliography{refs}

\clearpage
\appendix
\renewcommand{\thefigure}{A.\arabic{figure}}
\renewcommand{\thetable}{A.\arabic{table}}
\setcounter{figure}{0}
\setcounter{table}{0}
\captionsetup[figure]{name=Fig.}
\captionsetup[table]{name=Table}

\section{Supplementary data}\label{app:supp}

Supplementary material related to this article---including the 4,056-pair five-DOT benchmark, the page-disjoint train/dev/test splits, the held-out Michigan DOT transfer set, the 500 single-doc and 100 multi-plan CAD-generated compliance test sets, the 424-pair page-disjoint test split used for the main-body retrieval comparison (Table~\ref{tbl:comparison}) and the per-category VQA judge evaluation, the LoRA adapter checkpoints (head-gentle, head-standard, full-LM) and the training scripts, the complete set of visual grounding heatmaps, and the specific prompt templates used for each agent will be made available upon acceptance.

\section{Human validation of the generated datasets}\label{app:human-val}

Because both the QnA benchmark and the compliance test set are model-generated, we complement the automatic generate-then-verify pipeline with two human-grounded checks: the fully hand-curated anchor set described in Section~\ref{sec:eval-dataset}, and a large-scale structured web validation of the generated corpus.

\paragraph{Protocol.} All $4,656$ generated items---the $4,056$ QnA pairs (full five-DOT benchmark) together with the $500$ single-doc and $100$ multi-plan compliance drawings---are presented to human reviewers through a lightweight web interface. For each item the reviewer sees the original drawing at full resolution alongside the complete context needed for an independent judgement: for QnA pairs, the model-generated question and answer plus the source agency, plan identifier, sheet title, and reasoning category; for compliance drawings, the encoded design facts, the rule and threshold being checked, the named governing standard (e.g.,\ ACI~318, AASHTO, WYDOT~606.05, PROWAG/ADA), the model's compliant/non-compliant verdict, and the most relevant standard-plan page(s) retrieved by ColPali for that check. Each reviewer records one of three verdicts---\textsc{correct}, \textsc{incorrect}, or \textsc{unsure}---with optional free-text notes. The study runs in \emph{agreement mode}: every item is shown to every reviewer independently, so items reviewed by two or more people yield an inter-annotator agreement signal, and disagreements localize the pairs most worth re-examining. Per-reviewer verdicts are logged separately and aggregated post~hoc.

\paragraph{Status.} This larger-scale validation over all $4,656$ generated items is \emph{ongoing}: per-type confirmation rates, inter-annotator agreement on doubly-reviewed items, and flagged-item counts will be released together with the code and data. In the interim, the hand-curated 78-pair anchor set provides a fully human-grounded check, and the automatically generated-then-verified pipeline (an independent Qwen-VL-7B verifier that rejects roughly half of all drafted questions) guards label quality at scale; pending completion, we scope the benchmark's label-quality guarantee to the verified anchor set plus this automated two-stage filter.

\section{Bootstrap CIs on the 424-pair test split}\label{app:bench-stats}

This appendix reports bootstrap confidence intervals on the full 424-pair page-disjoint test split over the 1,898-page five-DOT visual index (no subset). Table~\ref{tbl:bench-stats-primary} gives per-category Recall@5 (generator-independent, clean zero-shot ColPali) and per-generator Judge Accuracy for the open generators (Qwen2.5-VL-7B, Qwen2.5-VL-72B, InternVL-2.5-8B), the latter scored by a local Qwen2.5-VL-72B judge; Table~\ref{tbl:bench-stats-retrieval} gives per-baseline, per-agency Recall@5. All CIs are percentile bootstrap with 10,000 resamples (seed~42).

\begin{table*}[!t]
\caption{Per-category Recall@5 (Hit) and Judge Accuracy (Judge) on the full 424-pair page-disjoint test split over the 1,898-page five-DOT index, with 95\% bootstrap CIs (10,000 resamples, seed~42). Recall@5 is a clean zero-shot ColPali and is generator-independent. Judge Accuracy is scored by a local Qwen2.5-VL-72B judge (no proprietary API). Per-category $N$: Dim$=$112, Vis$=$145, Log$=$120, Hal$=$47. Logical-Reasoning Recall@5 is markedly lower here (52.50\%) than in Table~\ref{tbl:category} (87.64\%) because this table uses the predecessor ColPali backbone, whose fixed $448{\times}448$ encoder is weakest on note/table-dense logical queries; the adopted ColNomic-3B closes this gap.}\label{tbl:bench-stats-primary}
\centering
\small
\begin{tabular*}{\textwidth}{@{\extracolsep{\fill}}llcc@{}}
\toprule
Model & Category & Hit (\%) & Judge (\%) \\
\midrule
  \multirow{5}{*}{Qwen2.5-VL-72B} & Dimensional  & 86.61 [80.4, 92.9] & 62.16 [53.2, 71.2] \\
   & Visual       & 86.21 [80.0, 91.7] & 78.47 [71.5, 84.7] \\
   & Logical      & 52.50 [43.3, 61.7] & 69.17 [60.8, 77.5] \\
   & Hallucination & 87.23 [76.6, 95.7] & 70.21 [57.4, 83.0] \\
  & \textbf{Overall} & \textbf{76.89 [72.9, 80.9]} & \textbf{70.62 [66.1, 74.9]} \\
\midrule
  \multirow{5}{*}{Qwen2.5-VL-7B} & Dimensional  & 86.61 [80.4, 92.9] & 60.36 [51.4, 69.4] \\
   & Visual       & 86.21 [80.0, 91.7] & 68.06 [60.4, 75.7] \\
   & Logical      & 52.50 [43.3, 61.7] & 59.17 [50.0, 68.3] \\
   & Hallucination & 87.23 [76.6, 95.7] & 57.45 [42.6, 70.2] \\
  & \textbf{Overall} & \textbf{76.89 [72.9, 80.9]} & \textbf{62.32 [57.6, 66.8]} \\
\midrule
  \multirow{5}{*}{InternVL-2.5-8B} & Dimensional  & 86.61 [80.4, 92.9] & 43.75 [34.8, 52.7] \\
   & Visual       & 86.21 [80.0, 91.7] & 60.00 [51.7, 68.3] \\
   & Logical      & 52.50 [43.3, 61.7] & 48.33 [39.2, 57.5] \\
   & Hallucination & 87.23 [76.6, 95.7] & 31.91 [19.1, 44.7] \\
  & \textbf{Overall} & \textbf{76.89 [72.9, 80.9]} & \textbf{49.29 [44.6, 54.0]} \\
\bottomrule
\end{tabular*}
\end{table*}

\begin{table*}[!t]
\caption{Per-baseline Recall@5 with 95\% bootstrap CIs on the 424-pair page-disjoint test split (10,000 resamples, seed 42). Per-agency $N$: WYDOT $=53$, Caltrans $=134$, AZDOT $=46$, CDOT $=31$, FDOT $=160$.}\label{tbl:bench-stats-retrieval}
\centering
\footnotesize
\setlength{\tabcolsep}{3pt}
\begin{tabular*}{\textwidth}{@{\extracolsep{\fill}}lcccccc@{}}
\toprule
Method & Overall (\%) & WYDOT & Caltrans & AZDOT & CDOT & FDOT \\
\midrule
ColQwen2.5-v0.2 & 87.26 [84.0, 90.3] & 92.5 [84.9, 98.1] & 87.3 [81.3, 92.5] & 82.6 [71.7, 93.5] & 87.1 [74.2, 96.8] & 86.9 [81.2, 91.9] \\
DSE-Qwen2-2B & 22.17 [18.2, 26.2] & 30.2 [18.9, 43.4] & 20.1 [13.4, 26.9] & 26.1 [13.0, 39.1] & 25.8 [9.7, 41.9] & 19.4 [13.1, 25.6] \\
VisRAG-Ret & 53.77 [49.1, 58.5] & 58.5 [45.3, 71.7] & 46.3 [38.1, 54.5] & 67.4 [54.3, 80.4] & 48.4 [32.3, 64.5] & 55.6 [47.5, 63.1] \\
BGE-M3 + OCR & 36.79 [32.3, 41.5] & 45.3 [32.1, 58.5] & 26.1 [18.7, 33.6] & 45.7 [32.6, 60.9] & 41.9 [25.8, 58.1] & 39.4 [31.9, 46.9] \\
CLIP ViT-B/32 & 1.89 [0.7, 3.3] & 0.0 [0.0, 0.0] & 0.0 [0.0, 0.0] & 8.7 [2.2, 17.4] & 0.0 [0.0, 0.0] & 2.5 [0.6, 5.0] \\
LayoutLMv3 & 0.00 [0.0, 0.0] & 0.0 [0.0, 0.0] & 0.0 [0.0, 0.0] & 0.0 [0.0, 0.0] & 0.0 [0.0, 0.0] & 0.0 [0.0, 0.0] \\
Nougat-decode + MiniLM & 0.47 [0.0, 1.2] & 1.9 [0.0, 5.7] & 0.0 [0.0, 0.0] & 0.0 [0.0, 0.0] & 0.0 [0.0, 0.0] & 0.6 [0.0, 1.9] \\
Pix2Struct-decode + MiniLM & 7.31 [5.0, 9.9] & 11.3 [3.8, 20.8] & 1.5 [0.0, 3.7] & 13.0 [4.3, 23.9] & 16.1 [3.2, 29.0] & 7.5 [3.8, 11.9] \\
UDOP-decode + MiniLM & 0.00 [0.0, 0.0] & 0.0 [0.0, 0.0] & 0.0 [0.0, 0.0] & 0.0 [0.0, 0.0] & 0.0 [0.0, 0.0] & 0.0 [0.0, 0.0] \\
OCR + MiniLM & 25.24 [21.2, 29.5] & 34.0 [20.8, 47.2] & 18.7 [12.7, 25.4] & 43.5 [30.4, 56.5] & 16.1 [3.2, 29.0] & 24.4 [17.5, 31.2] \\
VisionRAG (Pyramid, RRF) & 23.11 [19.1, 27.1] & 32.1 [18.9, 45.3] & 18.7 [11.9, 25.4] & 28.3 [15.2, 41.3] & 16.1 [3.2, 29.0] & 23.8 [17.5, 30.0] \\
HPC-ColPali (BQ) & 65.09 [60.6, 69.6] & 71.7 [58.5, 83.0] & 62.7 [54.5, 70.9] & 67.4 [54.3, 80.4] & 41.9 [25.8, 58.1] & 68.8 [61.2, 75.6] \\
\bottomrule
\end{tabular*}
\end{table*}

\section{Additional retrieval and judge analyses}\label{app:extra}

\paragraph{Rank-sensitive retrieval metrics.} Table~\ref{tbl:rankmetrics} reports Recall@1, Recall@5, MRR, and nDCG@10 for the adopted ColNomic-3B retriever on the 424-pair page-disjoint test split over the 1,898-page index. The $\approx$26-pp gap between Recall@5 ($92.45\%$) and Recall@1 ($66.75\%$) is material for the single-pass auditor, which consumes the rank-1 page; it motivates the optional cross-encoder re-ranker for rank-1-sensitive deployments.

\begin{table}[!t]
\caption{Rank-sensitive retrieval metrics for zero-shot ColNomic-3B on the 424-pair test split (1,898-page index).}\label{tbl:rankmetrics}
\centering\footnotesize
\setlength{\tabcolsep}{4pt}
\begin{tabular*}{\columnwidth}{@{\extracolsep{\fill}}lccccc@{}}
\toprule
Category & $N$ & R@1 & R@5 & MRR & nDCG@10 \\
\midrule
Dimensional   & 112 & 67.86 & 97.32 & 0.803 & 0.847 \\
Visual        & 145 & 68.28 & 93.10 & 0.788 & 0.828 \\
Logical       & 120 & 69.17 & 88.33 & 0.778 & 0.810 \\
Hallucination & ~47 & 53.19 & 89.36 & 0.682 & 0.741 \\
\midrule
\textbf{Overall} & \textbf{424} & \textbf{66.75} & \textbf{92.45} & \textbf{0.777} & \textbf{0.819} \\
\bottomrule
\end{tabular*}
\end{table}

\paragraph{Michigan transfer over the joint index.} To remove the candidate-pool-size confound, we re-evaluate Michigan retrieval over the \emph{combined} $1,898{+}298$-page index rather than the $298$-page Michigan-only pool (Table~\ref{tbl:michjoint}). Over the joint index Michigan transfers at $91.40\%$ Recall@5---essentially matching the in-distribution benchmark ($91.47\%$)---rather than the $93.55\%$ obtained on the smaller, easier Michigan-only pool. We therefore report the joint-index figure as the comparable transfer number and do not claim Michigan exceeds in-distribution agencies.

\begin{table}[!t]
\caption{Zero-shot Michigan transfer, joint vs.\ Michigan-only candidate pool.}\label{tbl:michjoint}
\centering\footnotesize
\setlength{\tabcolsep}{4pt}
\begin{tabular*}{\columnwidth}{@{\extracolsep{\fill}}lccc@{}}
\toprule
Candidate pool & \#\,Candidates & R@1 & R@5 \\
\midrule
Joint ($1,898{+}298$) & 2,196 & 68.82 & \textbf{91.40} \\
Michigan-only           & ~~298  & 74.19 & 93.55 \\
\bottomrule
\end{tabular*}
\end{table}

\paragraph{Cross-family judge agreement.} Because Qwen2.5-VL serves as drafter, verifier, answerer, and judge, we test for same-family self-preference by re-judging the 424 Qwen2.5-VL-7B zero-shot answers with a non-Qwen judge (InternVL2.5-8B) on identical inputs (gold page, question, reference, model answer). The two judges agree on $92.92\%$ of items with Cohen's $\kappa=0.75$ (substantial); the same-family Qwen judge is in fact \emph{stricter} ($79.95\%$ vs the cross-family $85.61\%$), the opposite of a self-preference bias. Judge verdicts are therefore not an artifact of evaluator--generator family overlap.

\section{Compliance pipeline confusion matrix and decomposition analysis}\label{app:compliance}

This appendix presents the full confusion matrix for the 8-query agentic compliance evaluation reported in Section~\ref{sec:res-agentic}. Table~\ref{tbl:compliance-matrix} gives the per-query predicted verdict, ground-truth verdict, target plans, and number of verification steps executed. Ground-truth verdicts are derived from the fact that the eight compliance queries target internally consistent WYDOT standard-plan configurations---no deliberately malformed inputs were constructed---so the correct verdict is PASS in every case.

\begin{table}[!t]
\caption{Agentic compliance confusion matrix over 8 cross-plan verification queries. GT and Pred are the ground-truth and predicted verdicts; Match indicates agreement; Steps is the number of verification steps the Planner produced. Expected minimum decomposition is 2 steps per query.}\label{tbl:compliance-matrix}
\centering
\resizebox{\columnwidth}{!}{%
\begin{tabular*}{1.15\columnwidth}{@{\extracolsep{\fill}}lllccr@{}}
\toprule
ID & Plans & GT & Pred & Match & Steps \\
\midrule
comp\_01 & 511-1A, 203-2A & PASS & PASS & \checkmark &  8 \\
comp\_02 & 606-7C, 606-2B & PASS & PASS & \checkmark &  7 \\
comp\_03 & 511-1A         & PASS & PASS & \checkmark &  8 \\
comp\_04 & 202-1          & PASS & PASS & \checkmark &  9 \\
comp\_05 & 511-1A         & PASS & PASS & \checkmark & 10 \\
comp\_06 & 606-4B         & PASS & PASS & \checkmark &  8 \\
comp\_07 & 203-2A         & PASS & PASS & \checkmark & 10 \\
comp\_08 & 203-2A         & PASS & PASS & \checkmark &  9 \\
\bottomrule
\end{tabular*}%
}
\end{table}

\paragraph{Over-decomposition.}
Expected minimum decomposition for each query is 2 steps (identify, then verify). The Planner produces 7--10 steps per query, averaging 8.6 steps overall, i.e., $\sim$4.3$\times$ the minimum. Over-decomposition does not compromise correctness in this evaluation because every expanded sub-step still retrieves a relevant plan region, but it does inflate end-to-end latency linearly in the step count.

\paragraph{Accuracy.}
$8/8 = 100.0$\%. The ground-truth construction (all-PASS) means this metric reflects the pipeline's ability to avoid false-positive rejections on valid plans; it is \emph{not} a direct measure of sensitivity to compliance violations. A complementary evaluation with deliberately malformed plans (insertion of inconsistent dimensions, missing notes, conflicting cross-references) is listed as follow-up work to characterise false-negative behavior, and is beyond the current scope because no such violation corpus exists in the WYDOT standard-plan archive.

\paragraph{Known limitations.}
(i)~The evidence-trail metric underreports actual grounding quality because it treats missing bounding-box metadata as missing evidence, even though the Auditor's rendered heatmaps and textual citations are present. A post-hoc extension to emit machine-parseable box coordinates would lift this metric without any change to the underlying reasoning. (ii)~The 8-query scale is too small for per-domain breakdowns; the eight queries are stratified across compliance domains (geotextile, guardrail, drainage, mailbox sight distance, gabion, silt fence, scarification, contour ditch) but each domain contributes only one query.

\subsection{False-Negative Sensitivity Study}\label{app:fn-sensitivity}

The 100\% accuracy reported in Section~\ref{sec:res-agentic} is measured on an all-PASS ground-truth set and therefore characterises the pipeline's \emph{false-positive rejection avoidance} rather than its \emph{false-negative violation-detection sensitivity}. Because the agentic pipeline is query-driven---the Auditor receives a retrieved standard-plan page from ColPali and compares any proposed design values \emph{stated in the query text} against the standard---we probe sensitivity by constructing a synthetic corpus of queries that describe deliberately non-compliant proposed designs, rather than by mutating the underlying plan images. This matches the natural deployment mode of the system (engineer submits a proposed value; the pipeline checks it against the relevant WYDOT standard) and isolates the Auditor's violation-detection behaviour from any retrieval-layer artefacts.

\paragraph{Violation taxonomy.}
Twenty-four synthetic-FAIL quer-ies are constructed, stratified evenly across three violation types (eight queries per type):
\begin{itemize}
    \item \textbf{Dimension mutation (M\textsubscript{dim})}: the query proposes a numeric value that violates the WYDOT specification for that component (e.g., ``A contractor proposed a steel stake vertical spacing of 15\,ft for silt fence installation per Plan 606-4B. Is this compliant with the required spacing?''---the true spacing is 4$'$-10$''$).
    \item \textbf{Note omission (M\textsubscript{note})}: the query describes a proposed construction procedure that skips a required note-level constraint (e.g., omitting the geotextile underlayment or the scarification step referenced in the grading notes).
    \item \textbf{Symbol / component swap (M\textsubscript{sym})}: the query substitutes a different physical component for one depicted in the plan (e.g., wooden stakes in place of steel stakes, or a concrete end anchor in place of a cable anchor).
\end{itemize}

Combined with the eight original PASS queries from Section~\ref{sec:res-agentic}, this yields an evaluation set of $N{=}32$ queries ($8$ PASS $+$ $24$ FAIL) that the pipeline processes blind. The full query list is provided in the released artefacts (\texttt{agentic\_compliance/run\_fn\_sensitivity. py}).

\paragraph{Detection rule.}
{\sloppy A query is counted as \emph{detected} if the Synthesizer's final verdict in the rendered Markdown report matches any of the regexes (case-insensitive): \texttt{FAIL}, \texttt{NON-COMPLIANT}, \texttt{DOES NOT} \texttt{(MEET|$|$}\allowbreak\texttt{COMPLY|$|$}\allowbreak\texttt{MATCH|$|$}\allowbreak\texttt{SATISFY)}, \texttt{VIOLAT}\allowbreak\texttt{(E|$|$ES|$|$ION)}, \texttt{EXCEEDS MAXIMUM}, or \texttt{BELOW MINIMUM}. This rule is conservative---it requires the pipeline to state the violation verdict explicitly rather than imply it.\par}

\paragraph{Metrics.}
We report the binary-classification confusion matrix together with the true-positive rate (TPR, equivalently \emph{sensitivity}), false-negative rate (FNR), specificity (TNR), precision, F1, and Cohen's~$\kappa$ against the synthetic ground truth:
\begin{equation}
\begin{aligned}
\text{TPR} &= \frac{\text{TP}}{\text{TP}+\text{FN}}, \quad \text{FNR} = \frac{\text{FN}}{\text{FN}+\text{TP}}, \\
\text{TNR} &= \frac{\text{TN}}{\text{TN}+\text{FP}}.
\end{aligned}
\end{equation}
Stratified TPR/FNR are additionally reported per violation type to identify which classes the Auditor is most likely to miss.

\begin{table}[!t]
\caption{False-negative sensitivity on the $N{=}32$ synthetic query corpus. Per-stratum verdict counts (PASS\,/\,FAIL); GT$=$FAIL rows quantify detection sensitivity, and the GT$=$PASS row repeats the Section~\ref{sec:res-agentic} PASS accuracy for reference.}
\label{tbl:fn-sensitivity}
\centering
\resizebox{0.8\columnwidth}{!}{%
\begin{tabular}{lcc}
\toprule
Stratum & $N$ & PASS / FAIL \\
\midrule
GT$=$PASS (originals)            & 8  & 8 / 0 \\
GT$=$FAIL, M\textsubscript{dim}  & 8  & 1 / 7 \\
GT$=$FAIL, M\textsubscript{note} & 8  & 1 / 7 \\
GT$=$FAIL, M\textsubscript{sym}  & 8  & 3 / 5 \\
\midrule
\textbf{Overall}                 & 32 & 13 / 19 \\
\bottomrule
\end{tabular}%
}
\end{table}

\paragraph{Results.}
{\sloppy The pipeline achieves perfect specificity (TNR\,=\,100\%) on the eight original PASS queries---zero false positives on valid designs---and an overall TPR of 79.17\% on the twenty-four FAIL queries, for a balanced-class $\kappa$ of 0.655.\par}
 The per-type breakdown is revealing: dimension-level (M\textsubscript{dim}) and note-omission (M\textsubscript{note}) violations are caught at identical $87.5$\% ($7/8$) sensitivity, whereas symbol/component-swap (M\textsubscript{sym}) violations fall to $62.5$\% ($5/8$). The three Msym\textsubscript{sym} misses are cases where the swapped component is visually plausible in the drawing (wooden vs.\ steel stakes, concrete vs.\ cable end anchor) and the Auditor's MaxSim retrieval surfaces the correct plan page, but the downstream VQA step does not explicitly contrast the proposed component against the one depicted, instead issuing a hedged PASS. Dimension mismatches---where the query states an explicit numeric value---trigger a numeric compliance check that the Auditor handles reliably. Precision is $100$\% by construction on this corpus because the pipeline never rejects a valid design on the PASS stratum, and the single overall metric most relevant for compliance deployment (FNR${=}20.83$\%) is dominated by the symbol-swap class.

\paragraph{Status.}
The query corpus, the runner script, the slurm submission wrapper, and the scoring harness (under \path{agentic_compliance/} and \path{evaluation/fn_sensitivity/}) are released alongside the primary evaluation artifacts. The numerical results in Table~\ref{tbl:fn-sensitivity} are measured values from job \texttt{53124015} on the \texttt{mb-a6000} partition (24/24 FAIL queries completed successfully, average pipeline latency $\approx 62$\,s per query). This study is cited as the immediate follow-up item in Section~\ref{sec:future}, and the symbol-swap failure mode is flagged there as the specific weakness that the next iteration should target (e.g., an explicit component-comparison sub-step after retrieval).

\section{Latency distribution}\label{app:latency}

This appendix reports end-to-end deployment latency. Table~\ref{tbl:efficiency} consolidates the one-time indexing cost and the per-query cost of each pipeline stage on the single-GPU target; Table~\ref{tbl:latency} gives the full per-query retrieval percentile distribution. All measurements are wall-clock time (batch size 1) on a single 80\,GB GPU (NVIDIA A100/H100).

\begin{table}[!t]
\caption{Retrieval-only latency distribution (ms) for full-page ColNomic MaxSim retrieval on the 424-pair page-disjoint test split over the combined 1,898-page five-DOT visual index. Wall-clock time collected with \texttt{time.perf\_counter()} on a single NVIDIA A100 80GB with PyTorch 2.x, one query at a time (batch size 1).}\label{tbl:latency}
\centering
\begin{tabular*}{\columnwidth}{@{\extracolsep{\fill}}lr@{}}
\toprule
Percentile & Full-page (424 queries, 1,898-page index) \\
\midrule
min  & 98.72 \\
p25  & 100.24 \\
p50  & 101.26 \\
p75  & 101.74 \\
p95  & 102.48 \\
p99  & 103.71 \\
max  & 138.19 \\
\midrule
mean & 101.24 \\
\bottomrule
\end{tabular*}
\end{table}

\begin{table}[!t]
\caption{Consolidated deployment latency on a single 80\,GB GPU (batch 1). Indexing is a one-time offline cost; all other rows are per query. The full retrieval percentile distribution is in Table~\ref{tbl:latency}; the agentic figure is the average over the compliance runs (Section~\ref{sec:res-agentic}).}\label{tbl:efficiency}
\centering\footnotesize
\setlength{\tabcolsep}{4pt}
\begin{tabular*}{\columnwidth}{@{\extracolsep{\fill}}lll@{}}
\toprule
Stage & Cost & Notes \\
\midrule
Indexing (1,898 pages) & $7.3$~min (one-time) & $4.34$~pages/s, offline \\
Retrieval / query        & $\approx$$0.10$~s (p50) & query enc.\ $+$ MaxSim \\
Re-rank (optional)       & $+$$\approx$$0.3$~s/cand.\ & $\le K$ passes; early-stop \\
VQA answer / query       & $2$--$5$~s & single-shot 72B \\
Agentic audit / query    & $60.9$~s & $\sim$$8.6$ sequential steps \\
\bottomrule
\end{tabular*}
\end{table}

\paragraph{Interpretation.}
The full-page retrieval latency distribution over the 1,898-page five-DOT index is tightly concentrated around the median (101~ms) with a heavy right tail (p99 $=$ 103.71~ms) driven by cold-cache query encoding on the first queries of each run. The mean of 101.24~ms/query reflects the linear corpus scaling implied by the MaxSim brute-force scoring loop. The tile-level variant trades higher per-query cost for finer spatial granularity; the binary-quantized HPC-ColPali variant (Section~\ref{sec:res-comparison}) adds only millisecond-scale unpacking overhead while compressing the index by an order of magnitude.

\paragraph{End-to-end latency.}
Retrieval is one component of end-to-end latency; VLM generation dominates the wall-clock budget. As a first-order estimate, Qwen-7B VQA on a $1024{\times}1024$ crop with 256 generation tokens takes 2--4 s on the same A100 80GB, so total interactive latency sits in the 2--5 s range per single-shot query, excluding the iterated Planner--Auditor--Synthesizer turns in the agentic compliance pipeline, which typically multiplies this by 3--5$\times$ depending on the number of plans cross-referenced.

\section{Grounding IoU pilot}\label{app:grounding}

This appendix summarizes the grounding IoU pilot study referenced in Section~\ref{sec:res-explainability}. The 16-query pilot set comprises: 12 queries drawn from the success/failure case studies in Section~\ref{sec:sample-cases} (steel-stake spacing, MDC depth, stake-spacing table, wire-tie location, plus-symbol convention, diaphragm identification, edge-drain trench width, V-mesh end-strip height, scarification depth, drainage fabric layout, contour ditch placement) and 4 additional dimensional-accuracy and visual-interpretation queries from the primary benchmark. Predicted bounding boxes are extracted from the sharpened MaxSim heatmap by thresholding at the top 5\% of patch activations and taking axis-aligned boxes over the resulting connected components. Ground-truth boxes are hand-drawn by a domain-informed reviewer at full plan resolution.

\paragraph{Metric definition.}
For each query, we compute the maximum IoU over all pairs of ground-truth and predicted boxes associated with that query (many-to-many max-IoU matching). Aggregate metrics include mean IoU, IoU$@0.3$ (fraction of queries with max IoU $\ge 0.3$), IoU$@0.5$, and IoU$@0.7$. IoU$@0.5$ is the primary headline metric for grounding evaluation, following common practice in object-detection pilots.

\paragraph{Pipeline status.}
The 16-query pipeline runs end-to-end: ColPali produces three candidate boxes per query from the sharpened MaxSim heatmap (48 predicted boxes in total), annotation-helper overlays are rendered for each query for reviewer inspection, and an IoU scorer computes many-to-many max-IoU against a ground-truth JSON. Pipeline artifacts---predicted boxes, annotation helpers, GT template, and scoring script---are released alongside the primary evaluation code. The pilot is deliberately small: the primary purpose is to establish that the MaxSim heatmap produces spatially-localized boxes that overlap meaningfully with human-identified answer regions, rather than to provide the sample size needed for per-category grounding claims.

\paragraph{Ground-truth annotation status.}
At the time of submission, the GT-box JSON was seeded with the predicted boxes so that the IoU pipeline could be exercised end-to-end; this configuration trivially yields IoU$=1.0$ on every query and is \emph{not} a meaningful measurement. Human GT annotation by a domain-informed reviewer is scoped as follow-up work, together with a larger-scale IoU study on a fully annotated corpus. We report this status explicitly rather than publishing a trivially perfect IoU headline number, because the downstream claim of interest---spatial localization quality---requires an honest GT and is not materially supported by the main results of this paper (which rely on Recall@5 and Judge Accuracy rather than pixel-level IoU). All 48 annotation-helper images are included in the supplementary bundle to assist with visually assessing localization quality prior to a formal IoU study.

\section{Sample case figures}\label{app:sample-cases}

The following figures show high-resolution screenshots of the ColPali + Qwen2.5-VL-7B Visual-First RAG pipeline applied to representative WYDOT benchmark queries. Each card displays the input query (top-left), the model's response (top-right), and the top-3 retrieved standard-plan pages with their MaxSim scores (bottom). Success or failure is determined by direct comparison of the model response to the benchmark ground-truth answer.

\subsection*{Success Cases (Judge Score = 1.0)}
In these cases the retriever surfaces the correct sheet and the VLM reads the queried dimension directly from the drawing, matching the ground truth (judge score 1.0).

\begin{figure*}[h]
  \centering
  \begin{subfigure}{\textwidth}
    \centering
    \includegraphics[width=\textwidth]{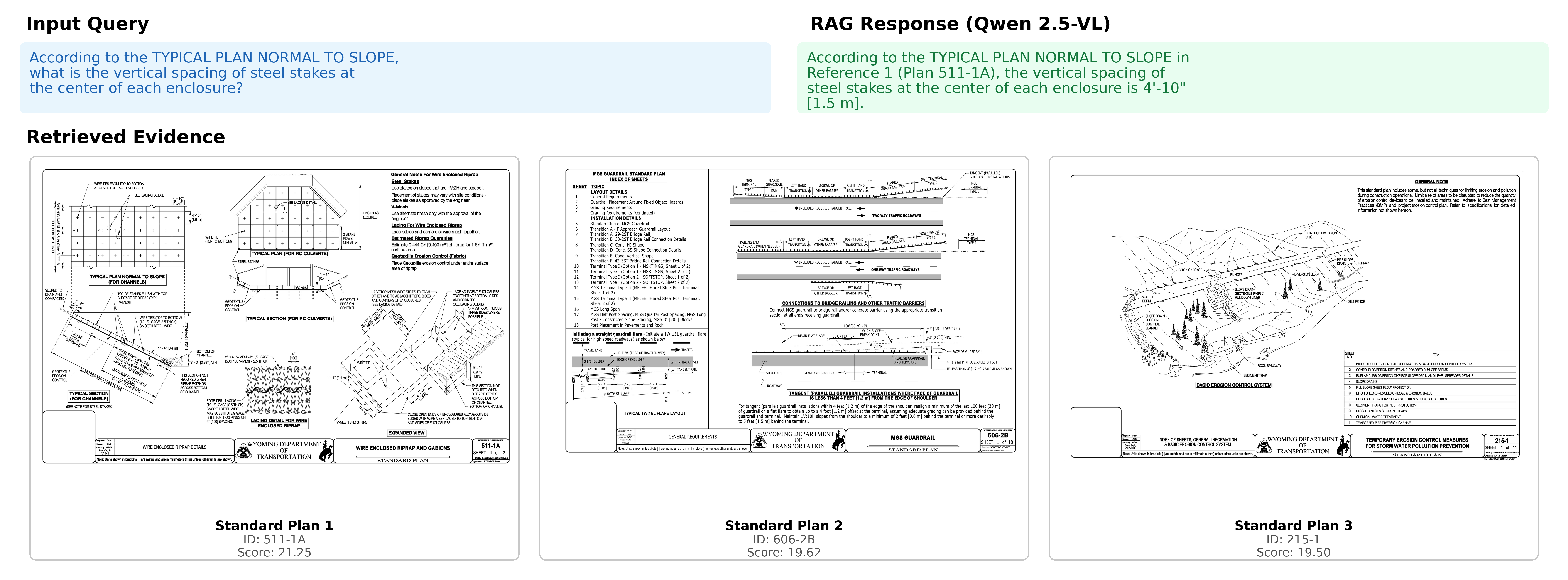}
    \caption{\textbf{Success} -- Dimensional accuracy. Q: vertical spacing of steel stakes at the centre of each enclosure (Plan 511-1A). Predicted ``4$'$-10$''$ [1.5~m]'' matches the ground truth.}
  \end{subfigure}
  \vskip 10pt
  \begin{subfigure}{\textwidth}
    \centering
    \includegraphics[width=\textwidth]{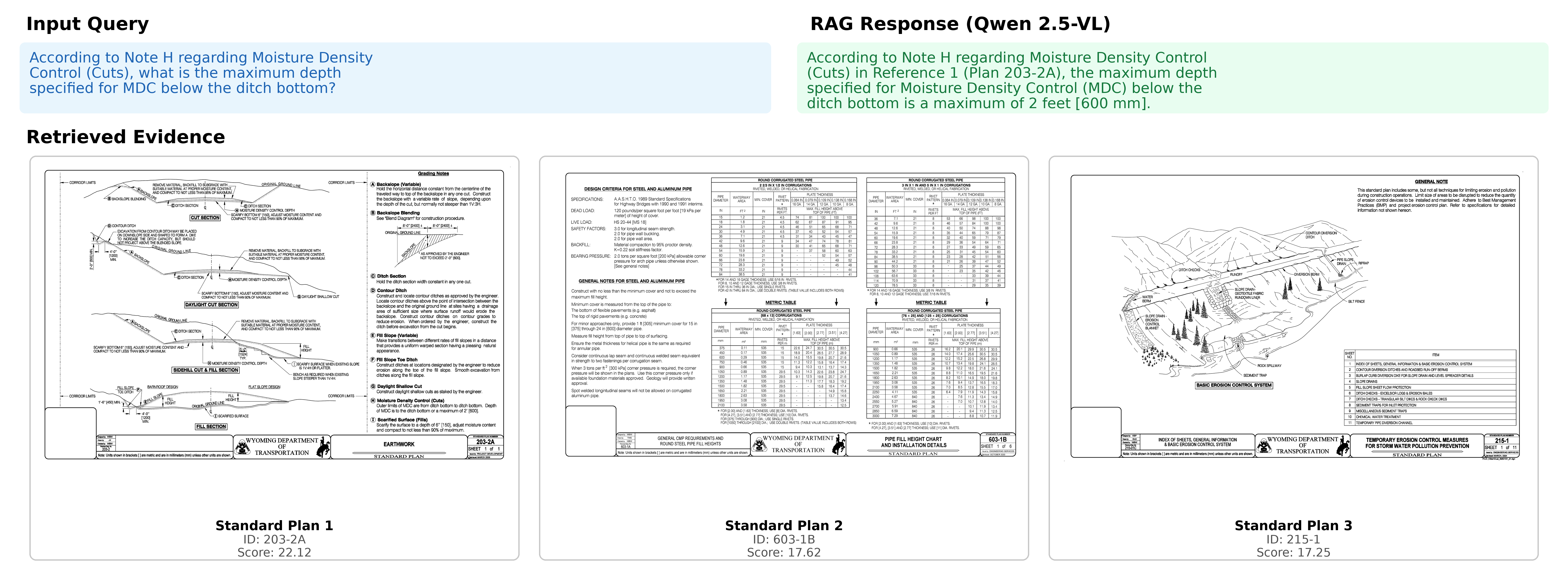}
    \caption{\textbf{Success} -- Dimensional accuracy. Q: maximum Moisture Density Control depth below the ditch bottom per Note H (Plan 203-2A). Predicted ``2~feet [600~mm]'' matches the ground truth.}
  \end{subfigure}
  \caption{Representative dimensional-accuracy success cases on the WYDOT benchmark.}
  \label{fig:success_cases_dim}
\end{figure*}

\subsection*{Failure Cases (Judge Score = 0.0)}
Here retrieval succeeds but the VLM binds the query to a neighboring dimension or conflates related quantities---the two dominant error modes analyzed in Section~\ref{sec:failure-analysis}.

\begin{figure*}[h]
  \centering
  \begin{subfigure}{\textwidth}
    \centering
    \includegraphics[width=\textwidth]{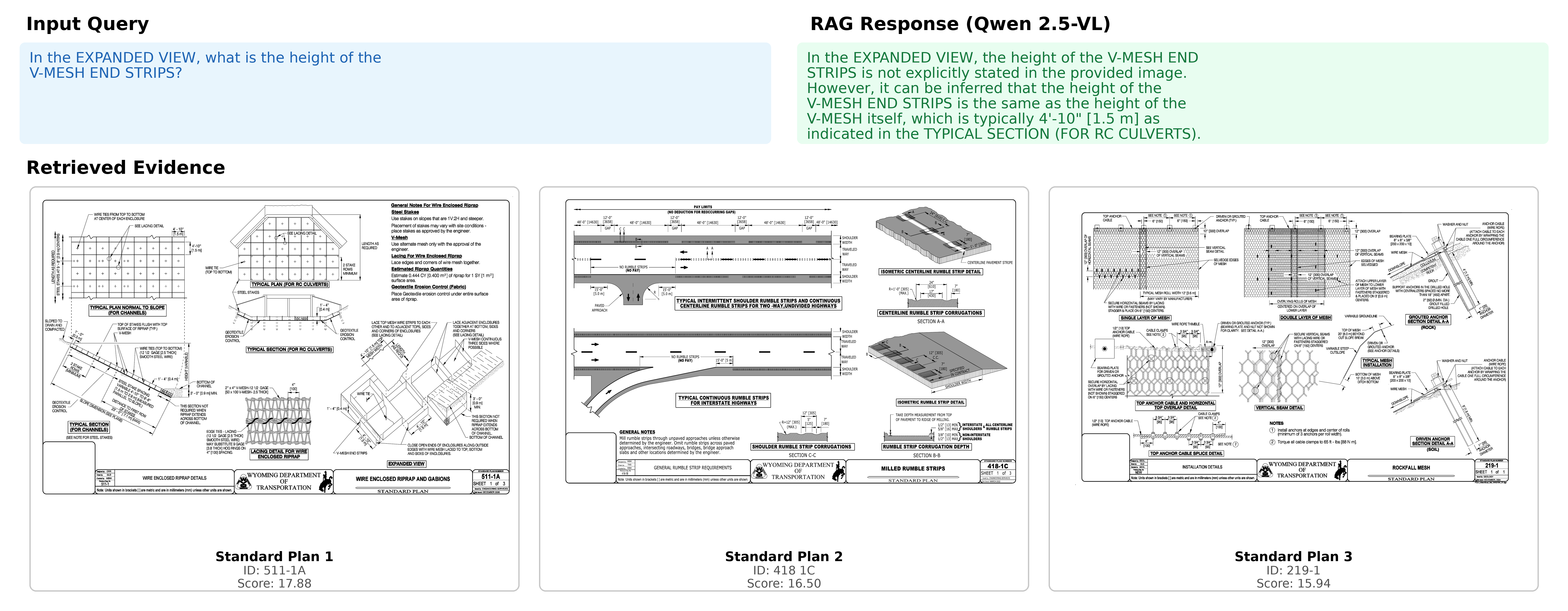}
    \caption{\textbf{Failure} -- Component--dimension binding. Q: height of the V-MESH END STRIPS in the EXPANDED VIEW (Plan 511-1A). Ground truth is 1$'$-4$''$ [0.4~m]; the model instead infers 4$'$-10$''$ [1.5~m] from a neighboring section, illustrating the cross-view inference error described in Section~\ref{sec:failure-analysis}.}
  \end{subfigure}
  \vskip 10pt
  \begin{subfigure}{\textwidth}
    \centering
    \includegraphics[width=\textwidth]{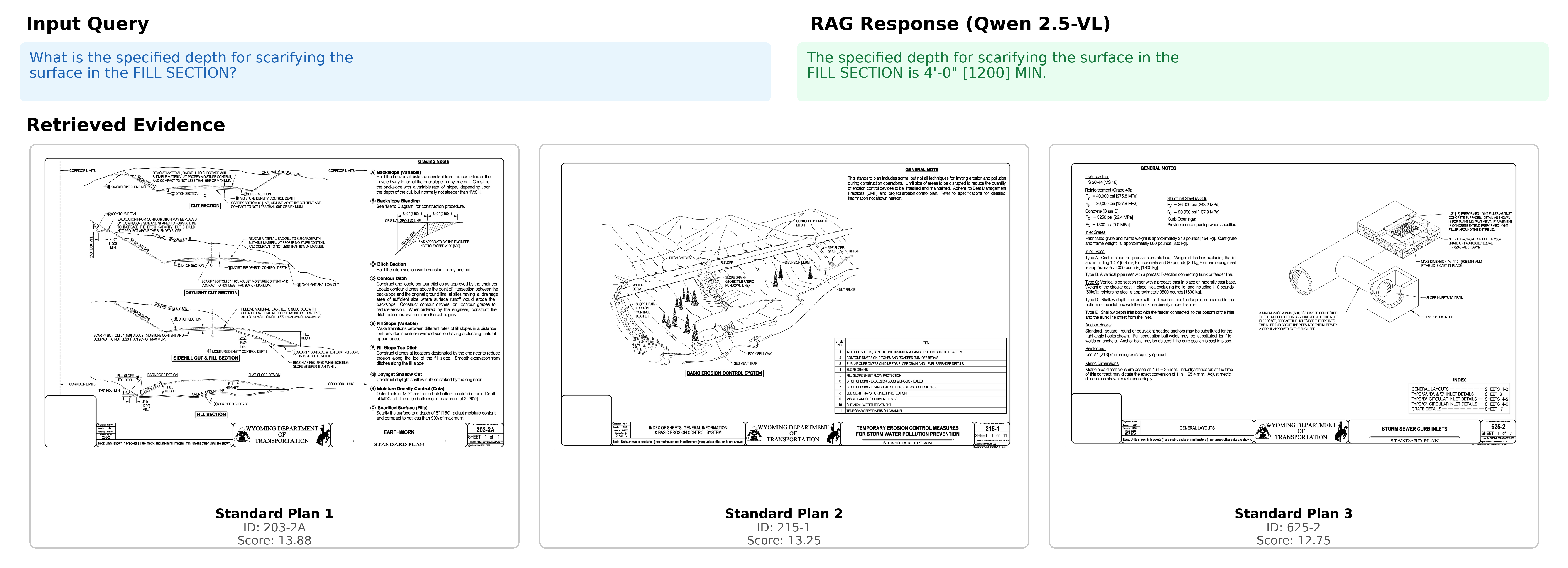}
    \caption{\textbf{Failure} -- Semantic misinterpretation. Q: specified depth for scarifying the surface in the FILL SECTION (Plan 203-2A). Ground truth is 6~inches [150~mm]; the model returns 4$'$-0$''$ [1200], a fill-geometry dimension, conflating ``scarification depth'' with ``fill thickness''.}
  \end{subfigure}
  \caption{Hard failures arising from component--dimension binding and semantic confusion.}
  \label{fig:failure_cases_dim}
\end{figure*}

\subsection*{Rule Grounding and Real-Plan Auditing Cases}
This is the input to the autonomous rule-grounding agent (Section~\ref{sec:res-rule-grounding}): the governing limit sits in one row among distractors, which the agent must retrieve and extract before auditing the design value.

\begin{figure*}[h]
\centering
\includegraphics[width=\columnwidth]{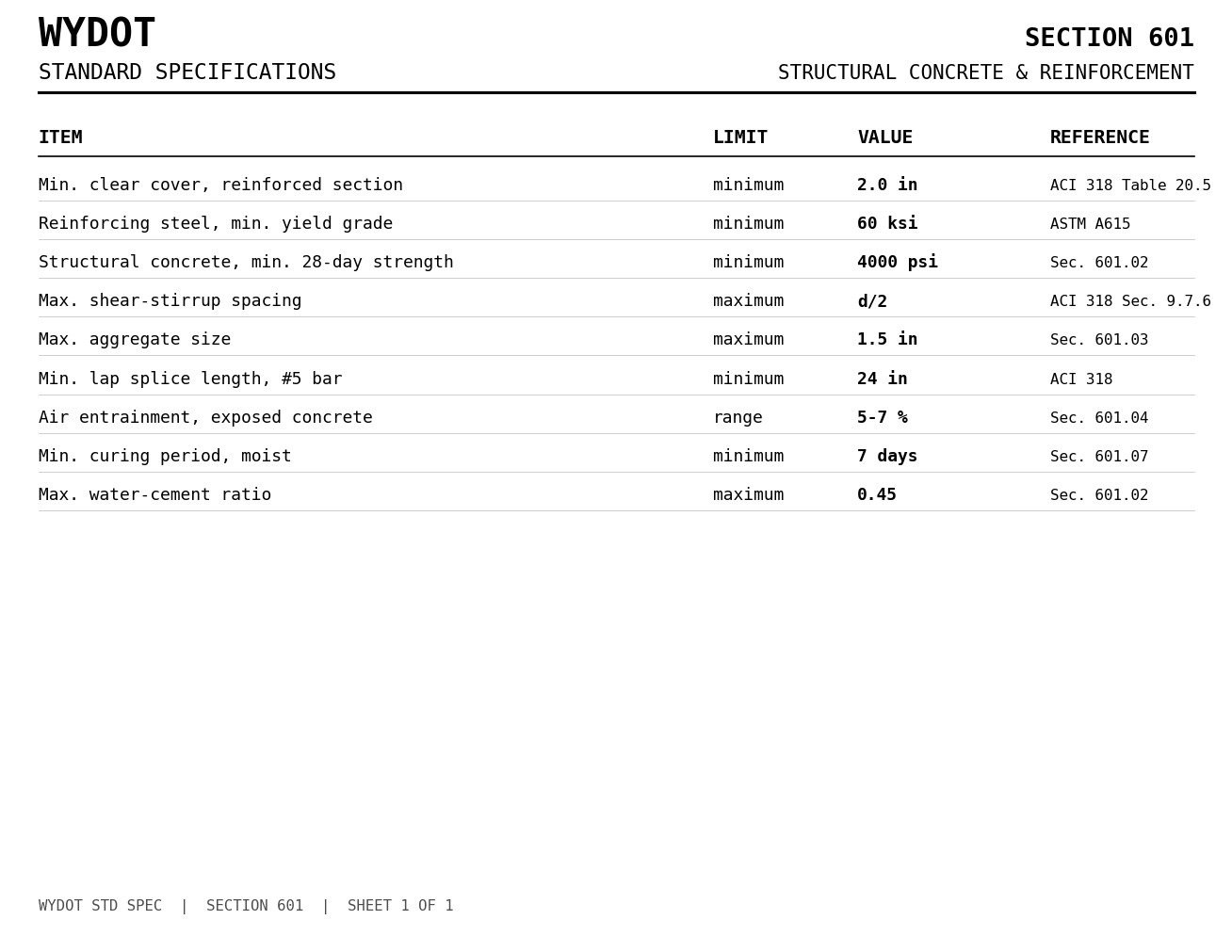}
\caption{A rendered Standard-Specifications reference sheet from the rule-grounding corpus. The seven governing limits for the compliance archetypes (e.g., min.\ cover $2.0''$, max.\ stirrup spacing $d/2$) are embedded among distractor rows; the agent must retrieve the correct sheet among 1,913 candidates and extract the governing limit from the table.}\label{fig:spec-sheet}
\end{figure*}

\clearpage 
\addtocounter{page}{-1}

\end{document}